\documentclass[11pt,letter]{article}
\usepackage[margin=1in]{geometry}
\usepackage{latexsym}
\usepackage{amsmath}
\usepackage{amssymb}
\usepackage{ifthen}
\usepackage{mathrsfs}
\usepackage{graphics}
\usepackage{color}
\usepackage{stmaryrd}
\usepackage{amsthm}

\usepackage{calc}
\usepackage[dvipsnames]{xcolor}
\usepackage{tikz}
\usepackage{tikz-cd}
\usetikzlibrary{calc,shapes,arrows}

\usepackage{hyperref}
\hypersetup{
  colorlinks=true,
  allcolors=black
}

\usetikzlibrary{arrows,positioning}
\tikzset{
    >=stealth',
    punkt/.style={
           rectangle,
           rounded corners,
           draw=black, very thick,
           text width=6.5em,
           minimum height=2em,
           text centered},
    pil/.style={
           ->,
           thick,
           shorten <=2pt,
           shorten >=2pt,}
}

\newcommand{\nc}{\newcommand}
\nc{\rnc}{\renewcommand} \nc{\nev}{\newenvironment}
\rnc{\subsection}{\secdef\ssa\ssb}
\nc{\ssa}[2][default]{\par\vspace{1ex}\refstepcounter{subsection}\noindent\textbf{\thesubsection.
#1. }} \nc{\ssb}[1]{\par\vspace{2ex}\noindent\textbf{#1. }}

\rnc{\subsubsection}{\secdef\sssa\sssb}
\nc{\sssa}[2][default]{\par\vspace{1ex}\refstepcounter{subsubsection}\noindent\textit{\thesubsubsection.
#1. }} \nc{\sssb}[1]{\par\vspace{1ex}\noindent\textit{#1. }}

\makeatletter
\rnc{\@seccntformat}[1]{{\normalfont\bfseries{\csname
the#1\endcsname}\hspace{1pt}.\hspace{0.4em}}}
\rnc{\section}{\@startsection
        {section}%
        {1}%
        {0mm}%
        {-\baselineskip}%
        {0.5\baselineskip}%
        {\normalfont\normalsize\bfseries\centering}%
}
\renewcommand{\@makecaption}[2]{\begin{center}#1. #2\end{center}}
\makeatother

\newtheorem{theo}{Theorem}[section]
\newtheorem{lem}[theo]{Lemma}
\newtheorem{cor}[theo]{Corollary}
\newtheorem{prop}[theo]{Proposition}

\theoremstyle{definition}
\newtheorem{defn}[theo]{Definition}
\newtheorem{rem}[theo]{Remark}

\newtheorem{exas}[theo]{Examples}

\rnc{\proof}[1][{}]{\smallskip\noindent\textit{Proof #1: }}
\nc{\proofend}{\hfill$\Box$\vspace{\topsep}\par}

\nc{\bende}{\eqno$\Box$}
\nc{\benda}{\tag*{$\Box$}}
\nc{\claimbenda}{\tag*{$\dashv$}}

\rnc{\labelenumi}{(\arabic{enumi})}
\rnc{\labelitemi}{\text{--}}
\rnc{\phi}{\varphi}
\rnc{\epsilon}{\varepsilon}
\nc{\bigmid}{\;\big|\;}
\nc{\Bigmid}{\;\Big|\;}
\rnc{\max}{\textup{max}}
\rnc{\min}{\textup{min}}
\rnc{\log}{\textup{log}\;}

\nc{\dotcup}{\;\dot\cup\;}
\nc{\dotbigcup}{\;\dot\bigcup\;}

\nc{\FO}{\textup{FO}}
\nc{\C}{\textup{C}}

\nc{\str}[1]{\ensuremath{\mathcal{#1}}}
\nc{\cls}[1]{\ensuremath{\mathsf{#1}}}
\nc{\Str}{\textsc{Str}}

\nc{\ind}{\textup{ind}}
\nc{\ord}{\textup{ord}}
\nc{\univ}{\textup{uni}}
\nc{\comp}{\textup{comp}}
\nc{\Mod}{\textsc{Mod}}
\nc{\Graph}{\textsc{Graph}}
\nc{\fin}{\textup{fin}}
\nc{\Forb}{\textsc{Forb}}

\newcommand{\Hom}{\textsc{Hom}}
\renewcommand{\hom}{\textup{hom}}
\nc{\Emb}{\textsc{Emb}}
\nc{\emb}{\textup{emb}}
\nc{\epi}{\textup{epi}}
\nc{\Epi}{\textsc{Epi}}
\nc{\aut}{\textup{aut}}
\nc{\Aut}{\textsc{Aut}}
\newcommand{\sHom}{\mbox{$s$-}\textsc{Hom}}
\newcommand{\shom}{\textup{s-hom}}
\nc{\sEmb}{\textsc{s-Emb}}
\nc{\semb}{\textup{s-emb}}
\nc{\sepi}{\textup{s-epi}}
\nc{\sEpi}{\textsc{s-Epi}}
\nc{\rank}{\textup{rank}}
\nc{\se}{\subseteq}

\newcommand{\vhom}{\ensuremath{\overrightarrow{\textup{hom}}}}

\newcommand{\subvec}[2]{\ensuremath{#1\!\!\upharpoonright_{#2}}}

\nc{\rand}[1]{\marginpar{\raggedright\footnotesize #1}}
\nc{\yrand}[1]{\rand{\textbf{Y: }#1}}
\nc{\jrand}[1]{\rand{\textbf{J: }#1}}
\nc{\zrand}[1]{\rand{\textbf{Z: }#1}}
\nc{\mrand}[1]{\rand{\textbf{M: }#1}}

\nc{\LV}{\textup{Lov\'{a}sz}}
\nc{\wi}{\textup{wi}}

\nc{\noP}{\ensuremath{\#\textup{P}}}

\begin{document}

\title{On Algorithms Based on Finitely Many Homomorphism Counts}

 \author{Yijia Chen\\\normalsize Department of Computer Science\\
 \normalsize Shanghai Jiao Tong University\\
 \normalsize yijia.chen@cs.sjtu.edu.cn\\
 \and
 J\"{o}rg Flum\\\normalsize Mathematisches Institut \\
 \normalsize Universit\"{a}t Freiburg i.Br.\\
 \normalsize flum@uni-freiburg.de
 \and
 Mingjun Liu\\\normalsize Department of Computer Science\\
 \normalsize Shanghai Jiao Tong University\\
 \normalsize liumingjun@sjtu.edu.cn\\
 \and
 Zhiyang Xun\\\normalsize Department of Computer Science\\
 \normalsize University of Texas at Austin\\
 \normalsize zxun@cs.utexas.edu\\
 }

\date{}
\maketitle

\begin{abstract}
It is well known~\cite{lov67} that up to isomorphism a graph~$G$ is
determined by the homomorphism counts $\hom(F, G)$, i.e., the number of
homomorphisms from $F$ to $G$, where $F$ ranges over all graphs.
Thus, in principle, we can answer any query concerning $G$ with only
accessing the $\hom(\cdot, G)$'s instead of $G$ itself. In this paper, we
deal with queries for which there is a \emph{hom algorithm}, i.e., there
are \emph{finitely many} graphs $F_1, \ldots, F_k$ such that for any graph
$G$ whether it is a \textsc{Yes}-instance of the query is already
determined by the vector
\[
\vhom_{F_1, \ldots, F_k}(G):= \big(\hom(F_1, G), \ldots, \hom(F_k, G)\big),
\]
where the graphs $F_1, \ldots, F_k$ only depend on $\varphi$.

We observe that planarity of graphs and 3-colorability of graphs,
properties expressible in monadic second-order logic, have no hom
algorithm. On the other hand, queries expressible as a Boolean combination
of universal sentences in first-order logic \FO\ have a hom algorithm. Even
though it is not easy to find \FO\ definable queries without a hom
algorithm, we succeed to show this for the non-existence of an isolated
vertex, a property expressible by the \FO\ sentence $\forall x\exists y
Exy$, somehow the ``simplest'' graph property not definable by a Boolean
combination of universal sentences. These results provide a
characterization of the prefix classes of first-order logic with the
property that each query definable by a sentence of the prefix class has a
hom algorithm.

For \emph{adaptive} query algorithms,
i.e., algorithms that again access $\vhom_{F_1, \ldots, F_k}(G)$ but here
$F_{i+1}$ might depend on $\hom(F_1, G), \ldots, \hom(F_i, G)$, we show
that \emph{three} homomorphism counts $\hom(\cdot, G)$ are both sufficient
and in general necessary to determine the isomorphism type of $G$. In
particular, by three adaptive queries we can answer any question on $G$.
Moreover, adaptively accessing two $\hom(\cdot, G)$'s is already enough to
detect the existence of an isolated vertex.

In~1993 Chaudhuri and Vardi~\cite{chavar93} showed the analogue of the \LV\
Isomorphism Theorem for the right homomorphism vector of a graph $G$, i.e,
the vector of values $\hom(G,F)$ where~$F$ ranges over all graphs
characterizes the isomorphism type of $G$. We study to what extent our
results carry over to the right homomorphism vector.
\end{abstract}

\section{Introduction}

In~\cite{lov67}, one of the first papers on graph homomorphisms, \LV\ proved
that graphs $G$ and $H$ are isomorphic if and only if for all graphs $F$ the
number $\hom(F,G)$ of homomorphisms from $F$ to~$G$ is equal to the number
$\hom(F,H)$ of homomorphisms from $F$ to $H$. Recently, this result has
attracted a lot of attention in various contexts, e.g., algorithms and
complexity~\cite{bokche19,curdelmar17}, machine learning~\cite{bok21,gro20a},
and logic~\cite{atskolwu21,gro20}. Among others, it provides a powerful
reduction of problems concerning graph structures to questions on the number
of homomorphisms, while homomorphisms have been the subject of extensive
study in the last few decades.

\LV' result says that the infinite vector
\[
\vhom(G):= \big(\hom(F,G)\big)_{\text{$F$ a graph}}
\]
determines the graph $G$ up to isomorphism. For a class $\cls C$ of graphs
set
\[
\vhom_{\cls C}(G):= \big(\hom(F,G)\big)_{F\in \cls C}.
\]
Using \LV' \emph{Cancellation Law}~\cite{lov71} (see
Theorem~\ref{thm:canclaw}) it is easy to see that for some classes~$\cls C$,
including the class of $3$-colorable graphs and the class of graphs that can
be mapped homomorphically to an odd cycle, $\vhom_{\cls C}(G)$ already
determines $G$ up to isomorphism. A further example: the class of
$2$-degenerate graphs has this property~\cite{dvor10}.

For other natural classes of graphs, $\vhom_{\cls C}(G)$ does not have the
full power of distinguishing non-isomorphic graphs but characterizes
interesting graph properties. For instance, let $\cls{TW}_k$ be the class of
graphs of tree-width bounded by $k$. It is shown in~\cite{delgrorat18} that
graphs $G$ and $H$ can be distinguished by the $k$-dimensional
Weisfeiler-Leman algorithm if and only if $\vhom_{\cls {TW}_k}(G) \ne
\vhom_{\cls {TW}_k}(H)$.

We turn to results more relevant for the algorithmic problems we are
interested in. Actually \LV' proof shows that in order to determine the
isomorphism type of $G$ it is sufficient to consider the homomorphism counts
$\hom(F,G)$ for the graphs $F$ with at most as many vertices as~$G$. As a
consequence, given an oracle to $\vhom(G)$,
we might deal with any query by first recovering the graph $G$ and then
considering the query on $G$. However, an algorithm based on this naive idea
requires \emph{exponentially} many entries in $\vhom(G)$, i.e., $\hom(F, G)$
for all isomorphism types of graphs~$F$ with $|V(F)|\le |V(G)|$, rendering
any practical implementation beyond reach.

There are queries that can be answered very easily using $\vhom(G)$, e.g., to
decide whether $G$ has a clique of size $k$, all we need to know is
$\hom(K_k, G)$ where $K_k$ is the complete graph on $k$ vertices. So ideally,
one would hope that to answer a query on $G$ it suffices to access a
\emph{constant} number of entries in $\vhom(G)$.

The question of using $\vhom(G)$ to answer queries algorithmically has been
raised before. In~\cite{curdelmar17} Curticapean et al.\ observed that
counting (induced) subgraphs isomorphic to a fixed graph $F$ can be reduced
to computing appropriate linear combinations of ``sub-vectors'' of
$\vhom(G)$. Thereby they introduced the so-called graph motif parameters.
Using this framework, they were able to design some algorithms to count
various specific subgraphs and induced subgraphs faster than the known ones.
These results can be understood as answering counting queries using
$\hom(\cdot, G)$'s. More explicitly, Grohe~\cite{gro20} asked whether it is
possible to answer any $\C^{k+1}$-query in \emph{polynomial time} by
accessing $\hom(F,G)$ for graphs $F$ of tree-width bounded by $k$. Here,
$\C^{k+1}$ denotes counting first-order logic with $k+1$
variables~\cite{caifurimm92}. Observe that without the polynomial time
constraint such an algorithm exists because graphs $G$ and $H$ cannot be
distinguished by $\C^{k+1}$ if and only if $\hom(F,G)= \hom(F,H)$ for
finitely many graphs $F$ of tree-width bounded by $k$~\cite{dvor10}(see
also~\cite{delgrorat18}).

\subsection*{Our contributions}
In this paper we study what Boolean queries (equivalently, graph properties)
can be answered using a \emph{constant number} of homomorphism counts. More
precisely, let $\cls C$ be a class of graphs closed under isomorphism. We ask
whether there are graphs $F_1, \ldots, F_k$ such that for any graph~$G$
whether $G\in \cls C$ can be decided by the finite vector
\[
\vhom_{F_1, \ldots, F_k}(G):= \big(\hom(F_1, G), \ldots, \hom(F_k, G)\big).
\]
In Section~\ref{sec:uni} we observe that this is the case if $\cls C$ can be
defined by a sentence of first-order logic (\FO) that is a Boolean
combination of \emph{universal} sentences. For $d\ge 1$ this includes the
class of graphs of maximum degree~$d$, of tree-depth~\cite{cheflu18} exactly
$d$, and the class of graphs of SC-depth~\cite{cheflu20} exactly~$d$ (but
also the classes where we replace ``exactly $d$'' by ``at most $d$''). On the
negative side, in Section~\ref{sec:isol} we show that for any $k\ge 1$ and
any $F_1, \ldots, F_k$ there are graphs $G$ and $H$ such that
\begin{itemize}
\item $\vhom_{F_1, \ldots, F_k}(G)= \vhom_{F_1, \ldots, F_k}(H)$ and $G$
    contains an isolated vertex but $H$ does not.
\end{itemize}
That is, any $\vhom_{F_1, \ldots, F_k}(\cdot)$ is not sufficient to detect
the existence of an isolated vertex. This is our technically most challenging
result, which requires some non-trivial argument using linear algebra. We
introduce some of the tools and construction methods needed in this proof
already in Section~\ref{sec:plan}, thereby showing the corresponding result
for the class of planar graphs and the class of 3-colorable graphs.

A graph $G$ has no isolated vertex if and only if it satisfies the
\FO-sentence $\forall x\exists y Exy$.
Thus we know for what quantifier-prefix classes of \FO-sentences all queries
definable by a sentence of the class can be answered by $\vhom_{F_1, \ldots,
F_k}(G)$ for some $F_1, \ldots, F_k$ \emph{independent} of~$G$.

Answering a query using $\vhom_{F_1, \ldots, F_k}(\cdot)$ can be phrased as
an algorithm checking this query with \emph{non-adaptive} access to the
vector $\vhom(G)$ on entries $F_1, \ldots, F_k$. It is also very natural to
allow access to $\vhom(G)$ to be \emph{adaptive}. Informally, on input $G$ an
adaptive algorithm still queries some $\hom(F_1, G), \ldots, \hom(F_k, G)$,
but for $i= 0, \ldots, k-1$ the choice of $F_{i+1}$ might depend on
$\hom(F_1, G), \ldots, \hom(F_i, G)$ (see~Definition~\ref{def:ada} for the
precise notion). It turns out that adaptive query algorithms are extremely
powerful. We first present an adaptive algorithm with \emph{two} accesses\ to
$\vhom(G)$ that can decide whether~$G$ contains an isolated vertex (see
Section~\ref{sec:star}). Even more, the algorithm is able to compute all the
information on the degrees of vertices in $G$. So in particular, it can
decide whether $G$ is regular. In Section~\ref{sec:ada} we provide an
adaptive algorithm that queries \emph{three} entries in $\vhom(G)$ that
determines (the isomorphism type of) $G$. Hence, it can answer any question
on $G$. The downside of this algorithm is its superpolynomial running time,
while all the aforementioned query algorithms run in polynomial time. We
conjecture that there is no polynomial time algorithm that can reconstruct an
input graph $G$ with access to $\vhom(G)$ (even without the requirement of a
constant number of accesses).

Results in~\cite{atskolwu21} may be interpreted as saying that often proper
sub-vectors of the \emph{right} homomorphism vector
\[
(\hom(G,F))_{\text{$F$ a graph}}
\]
of a graph $G$ are not so expressive as the corresponding sub-vectors of the
(left) homomorphism vector. For our topic, the finite sub-vectors, in
Section~\ref{sec:right} we prove that our two ``positive results'' on hom
algorithms (namely, Theorem~\ref{thm:exiuni} on Boolean combinations of
universal sentences and Theorem~\ref{thm:3adapiso} on the power of 3 adaptive
hom algorithms) fail for the finite right sub-vectors. Even when the result
is the same (e.g., there is no right hom algorithm showing the non-existence
of isolated vertices)
the proof and its complexity can be quite different.

\section{Preliminaries}\label{sec:prel}

We denote by $\mathbb N$ the set of natural numbers greater than or equal to
0. For $n\in \mathbb N$ let $[n]:= \{1, 2, \ldots, n\}$.

For graphs we use the notation $G= (V(G), E(G))$ common in graph theory. Here
$V(G)$ is the non-empty set of vertices of $G$ and $E(G)$ is the set of
edges. We only consider finite, simple and undirected graphs and briefly
speak of graphs. To express that there is an edge connecting the vertices $u$
and $v$ of the graph $G$, we use (depending on the context) one of the
notations $uv\in E(G)$ and $\{u,v\}\in E(G)$. For graphs $G$ and $H$ with
disjoint vertex sets we denote by $G\dotcup H$ the \emph{disjoint union of
$G$ and $H$}, i.e., the graph with vertex set $V(G)\cup V(H)$ and edge set
$E(G)\cup E(H)$. If the vertex sets are not disjoint, we tacitly pass to
isomorphic copies with disjoint vertex sets. Similarly, $\dotbigcup_{i\in
I}G_i$ denotes the disjoint union of the graphs $G_i$ with $i\in I$. For a
graph $G$ and $\ell\ge 1$ we denote by $\ell G$ the disjoint union of $\ell$
copies of $G$.

For $n\ge 1$ we denote by $K_n$ a clique with $n$ vertices, by $P_n$ a path
of $n$ vertices, and for $n\ge 3$ by $C_n$ a cycle of $n$ vertices.

For graphs $G$ and $H$ by $G\cong H$ we express that $G$ and $H$ are
isomorphic. All classes of graphs considered in this paper are closed under
isomorphism.

\begin{defn}\label{def:homHom}
Let $G$ and $H$ be graphs and $f:V(G)\to V(H)$. The function $f$ is a
\emph{homomorphism} if $uv\in E(G)$ implies $f(u)f(v)\in E(H)$ for all
$u,v\in V(G)$. It is an \emph{embedding} if $f$ is a homomorphism that is
one-to-one. We call $f$ an \emph{epimorphism} if $f$ is a homomorphism, the
range of $f$ is $V(H)$, and for every $u'v'\in E(H)$ there are $u,v\in V(G)$
with $uv\in E(G)$ and with $f(u)=u'$, $f(v)=v'$. We get the definitions of
\emph{strong homomorphism}, of \emph{strong embedding}, and of \emph{strong
epimorphism} by additionally requiring in the previous definitions that
\big($uv\in E(G) \iff f(u)f(v)\in E(H)$\big) for all $u,v\in V(G)$.

We denote by $\Hom(G,H)$ the set of homomorphisms from $G$ to $H$. Thus
$\hom(G,H):=|\Hom(G,H)|$ is the number of homomorphisms from~$G$ to $H$.
Similarly, we define $\sHom(G,H)$ and $\shom(G,H)$ for strong homomorphisms
and use corresponding notations for the other notions of morphisms. Finally,
$\Aut(G)$ and $\aut(G)$ denote the set of automorphisms of~$G$ and its
number, respectively.
\end{defn}

\noindent For example,
\begin{eqnarray}\label{eq:veredg}
\hom(K_1,G)=|V(G)| & \text{and} & \hom(P_2,G)=2\cdot |E(G)|.
\end{eqnarray}
The following equalities can easily been verified and will often tacitly been
used later. For graphs $F_1$, $F_2$, and $G$,
\begin{equation}\label{eq:unionhom}
\hom(F_1\dotcup F_2,G)= \hom(F_1,G)\cdot \hom(F_2,G),
\end{equation}
\begin{equation}\label{eq:unionemb}
\text{if $G$ is a connected graph,
 then $\emb(G,F_1\dotcup F_2)=\emb(G,F_1)+ \emb(G, F_2)$}.
\end{equation}

\noindent
For graphs $F$ and $G$
by $F\le G$ we mean that
\begin{equation}\label{eq:basord}
|V(F)|< |V(G)| \quad \text{ or} \quad \big(|V(F)|= |V(G)|
	\ \text{ and\ } \ |E(F)|\le |E(G)|\big).
\end{equation}
Note that there are non-isomorphic graphs $F$ and $G$ with $F\le G$ and $G\le
F$.
%
\begin{theo}[Lov\'asz Isomorphism
Theorem~\cite{lov67}]\label{thm:loviso} Let $G$ and $H$ be graphs. If
$\hom(F,G)= \hom(F,H)$ for all graphs $F$ with $|V(F)|\le \min \{|V(G)|,$
$|V(H)|\}$, then $G$ and $H$ are isomorphic. Hence, the finite vector
\[
\vhom(G)_{\{F \ \mid \ |V(F)|\le |V(G)|\}}
\]
determines $G$ up to isomorphism.
\end{theo}

\section{Algorithms accessing morphism counts}\label{sec:quealg}

For what classes $\cls C$ of graphs is there a finite set $\cls F$ of graphs
such that the membership of any graph~$G$ in $\cls C$ is determined by the
values $\hom(F,G)$, where $F$ ranges over $\cls F$? This question leads to
the following definition.

\begin{defn}\label{def:queryalg}
Let $\cls C$ be a class of graphs. \emph{A hom algorithm for $\cls C$
(with a constant number of non-adaptive accesses to homomorphism counts)}
consists of a $k\ge 1$, graphs $F_1,\ldots, F_k$, and a $X\subseteq \mathbb
N^k$ such that for all~$G$,
\[
G\in\cls C \iff \big(\hom(F_1,G),\ldots, \hom(F_k,G)\big)\in X.
\]
We then say that the \emph{hom algorithm decides $\cls C$}. Analogously
we define the notions of \emph{emb algorithm}, \emph{s-hom
algorithm}, and \emph{s-emb algorithm}.
\end{defn}

We often will use the following fact whose proof is immediate: A class~$\cls
C$ can be decided by a hom algorithm if and only if there are finitely many
graphs $F_1,\ldots, F_k$ such that for all $G$ and $H$ \big(recall that
$\vhom_{F_1,\ldots, F_k}(G)=(\hom(F_1,G),\ldots, \hom(F_k,G))$\big),
\[
\text{$\vhom_{F_1,\ldots, F_k}(G)=
	\vhom_{F_1,\ldots, F_k}(H)$ \ \ implies \ \ \big($G\in\cls C\iff H\in\cls C$\big).}
\]

\begin{rem}
If the set $X$ in Definition~\ref{def:queryalg} is decidable, then we easily
extract an actual algorithm with an oracle to $\vhom(G)$. However the above
equivalence only holds for arbitrary $X$. Nevertheless, all our positive
results have decidable $X$'s. We use the current definition to ease
presentation, and also to make our negative result, e.g.,
Theorem~\ref{thm:isol}, stronger.
\end{rem}
\begin{rem}\label{apprem:xdxcd}
Let $\cls C$ have a hom algorithm as
Definition~\ref{def:queryalg}, i.e., there are $k\ge 1$, graphs $F_1,
\ldots, F_k$ and a set $X\subseteq \mathbb N^k$ with
\begin{equation*}
G\in\cls C \iff \vhom_{F_1,\ldots, F_k}(G)\in X.
\end{equation*}
Then the set $X$ can be chosen to be decidable if and only if $\cls C$ is
decidable.

In fact, we set $X(\cls C):=\{\big(\hom(F_1,G)\ldots, \hom(F_k,G)\big) \mid
G\in\cls C \}$ and show that ($\cls C$ is decidable if and only if $X(\cls
C)$ is decidable). The direction from right to left is trivial. We turn to
the other direction. The main tool is the following trivial observation:
\begin{quote}
\emph{Let $F$ and $G$ be graphs and $h: V(F)\to V(G)$ a function.
Furthermore, let $G'$ be an induced subgraph containing the image of $V(F)$
under $h$. Then
\[
h\in\hom(F,G)\iff h\in\hom(F,G').
\]}
\end{quote}
Hence, a tuple $(\ell_1,\ldots, \ell_k)$ is in $X(\cls C)$ if and only if
there is a graph $G\in \cls C$ with $|V(G)|\le 1+ \ell_1\cdot |V(F_1)|+\cdots
+ \ell_k\cdot |V(F_k)|.$ Hence, if $\cls C$ is decidable, then so is $X(\cls
C)$ (note that for $\ell_i=0$ for all $i\in[k]$ we can take as $G$ a graph
with a single vertex).
\end{rem}

\begin{exas}\label{exa:basic}
(a) By taking $k=1$, a graph $F$ whose vertex set is a singleton, and $X:=
\{2n+1 \mid n\ge 1 \}$, we get a hom algorithm for the class of graphs with
an odd number of vertices (by~\eqref{eq:veredg}).
	
\smallskip
\noindent (b) \LV\ Isomorphism Theorem (Theorem~\ref{thm:loviso}) shows that
every class that only contains finitely many graphs up to isomorphism can be
decided by a hom algorithm.
	
\smallskip
\noindent (c) By passing from $k\ge 1$, $F_1,\ldots, F_k$, and $X\subseteq
\mathbb N^k$ to $k\ge 1$, $F_1,\ldots, F_k$, and $\mathbb N^k\setminus X$, we
see that with every class $\cls C$ also the class $\cls C^{\textup{comp}}:=
\{G \mid G\notin \cls C \}$ has a hom algorithm.
\end{exas}
\noindent By Definition~\ref{def:queryalg} we have four types of algorithms
(\hom, \emb, s-\hom, s-\emb). The following proposition shows that a class
has an algorithm of one type if and only if it has an algorithm of any other
type. This allows us to speak of a \emph{query algorithm accessing morphism
counts} (or \emph{query algorithm} for short) if in the given context it is
irrelevant to what type we refer.

\begin{prop}\label{pro:proeq}
For a class $\cls C$ of graphs the following are equivalent.
\begin{enumerate}\vspace{-2mm}
\item[(i)] There is a hom algorithm for $\cls C$.
		
\vspace{-2mm}
\item[(ii)] There is an emb algorithm for $\cls C$.
	
\vspace{-2mm}
\item[(iii)] There is an s-hom algorithm for $\cls C$.

\vspace{-2mm}
\item[(iv)] There is an s-emb algorithm $\cls C$.

\vspace{-2mm}
\item[(v)] There is a hom algorithm for $\cls C^c$, the class of
	graphs that are complements of graphs in~$\cls C$ (the complement of a
	graph $G$ is the graph $G^c= \big(V(G), \{uv \mid u\ne v \text{ and\ }
	uv\notin E(G)\}\big)$.
\end{enumerate}
\end{prop}

\proof The equivalences between the statements (i), (ii), (iii), and (iv) are
well known (see~\cite{gro20,lov67,lov12}). Among others they use the fact
that every $h\in \Hom(F,G)$ can be written as $h = f\circ g$, where for some
graph $F'$ we have $g\in \Epi(F,F')$ and $f\in\Emb(F',G)$. Clearly, $F'\le F$
(as otherwise $\Epi(F,F')= \emptyset$). Hence.
\begin{equation}\label{eq:homepiemb}
\hom(F,G)= \sum_{F'\le F}\frac{1}{\aut(F')}
	\cdot \epi(F,F')\cdot \emb(F',G),
\end{equation}
where the sum ranges over all isomorphism types of graphs~$F'$ with $F'\le
F$.

The equivalence between (iv) and (v) immediately follows from the equality
$\sEmb(F, G)= \sEmb(F^c, G^c)$. \proofend

\begin{rem}\label{rem:typechange}
The proofs of the equivalences of the previous proposition show the
following: If for $\cls C$ we have a query algorithm of one type based on
graphs $F_1, \ldots, F_k$ and $m:= \max \big\{|V(F_i)| \bigmid i\in
[k]\big\}$, then for any other type we can compute finitely many graphs, all
with at most $m$ vertices, that are the graphs of a query algorithm for $\cls
C$ of this other type.
\end{rem}

\section{FO-definable classes with query algorithms}\label{sec:uni}

We start by showing that every class of graphs definable by a (finite) set of
forbidden induced subgraphs has a query algorithm. Of course, the complement
and the union of such classes again have such an algorithm. In terms of
first-order logic this means that every class axiomatizable by a Boolean
combination of universal sentences has a query algorithm.
\medskip

\noindent
Let $\cls F$ be a finite set of graphs. We set\vspace{-1mm}
\[
\textsc{Forb}(\cls F)
:= \{G \mid \text{no induced subgraph of $G$ is isomorphic to a graph
in $\cls F$}\}.\vspace{-1mm}
\]
We say that \emph{a class $\cls C$ of graphs is definable by a set of
forbidden induced subgraphs} if there is a \emph{finite} set $\cls F$ with $\cls
C= \textsc{Forb}(\cls F) $.

If $\cls F:= \{F\}$ with $V(F):=[2]$ and $E(F):= \{12\}$, then
$\textsc{Forb}(\cls F)$ is the class of graphs without edges. Examples of
classes definable by a set of forbidden induced subgraphs are classes of
bounded vertex cover number (attributed to \LV), of bounded
tree-depth~\cite{din92}, or even of bounded shrub-depth~\cite{ghnoor12}. All these classes have a query algorithm:

\begin{lem}\label{lem:forb}
	Every class of graphs definable by a set of forbidden induced subgraphs can
	be decided by a query algorithm.
\end{lem}

\proof If $\cls C=\textsc{Forb}(\emptyset)$, we set $k= 1$, let $F$ be an
arbitrary graph, and take $X:= \mathbb N$. Assume now that $\cls C=
\textsc{Forb}(\cls F)$ with $\cls F= \{F_1,\ldots, F_k\}$ and $k\ge 1$. Then
\[
G\in \cls C \ \ \iff \ \ \semb(F_1,G)=\ldots =\semb(F_k,G)=0.
\]
Hence, $k$, $F_1,\ldots, F_k$, and $X\subseteq\mathbb N^k$ with $X= \big\{(0,
0, \ldots, 0)\big\}$ constitute an $\semb$-algorithm for $\cls C$. \proofend

The following lemma shows that the universe of classes with query algorithms
is closed under the Boolean operations. Part (a) was already mentioned as
part (c) of Examples~\ref{exa:basic}. We omit the straightforward proof.
\begin{lem}\label{lem:boole}\begin{enumerate}\vspace{-1mm}
		\item[(a)] If $\cls C$ has a query algorithm, then so does $\{G \mid G\notin
		\cls C\}$.
		
		\vspace{-2mm}
		\item[(b)] If $\cls C$ and $\cls C'$ have query algorithms, then $\cls
		C\cap \cls C'$ and $\cls C\cup \cls C'$ have query algorithms.
	\end{enumerate}
\end{lem}

\noindent
Recall that \emph{formulas} $\varphi$ of \emph{first-order logic} \FO\ for
graphs are built up from \emph{atomic formulas} $x_1= x_2$ and $Ex_1x_2$
(where $x_1, x_2$ are variables) using the Boolean connectives $\neg$,
$\wedge$, and $\vee$ and the universal~$\forall$ and existential $\exists$
quantifiers. A \emph{sentence} is a formula without free variables (i.e., all
variables of $\varphi$ are in the scope of a corresponding quantifier). If
$\varphi$ is a sentence, then we denote by $\cls C(\varphi)$ the class of
graphs that are models of $\varphi$.

An \FO-formula is \emph{universal} if it is built up from atomic and negated
atomic formulas by means of the connectives~$\wedge$ and~$\vee$ and the
universal quantifier $\forall$.
If in this definition we replace the universal quantifier
by the existential one, we get the definition of an \emph{existential
	formula}. The following result is well known (e.g., see~\cite{mck78}).

\begin{lem}\label{lem:exuni}
Let $\cls C$ be a class of graphs. Then\vspace{-1mm}
\begin{align*}
\text{$\cls C=\cls C(\varphi)$ for some universal $\varphi$}
\iff \text{$\cls C$ is definable by a set of forbidden induced subgraphs}.
\end{align*}
\end{lem}


\noindent
By Lemma~\ref{lem:forb} -- Lemma~\ref{lem:exuni} we get:
\begin{theo}\label{thm:exiuni}
If the \FO-sentence $\varphi$ is a Boolean combination of universal
sentences, then there is a query algorithm for $\cls C(\varphi)$.
\end{theo}


\begin{rem}
The class $\cls C(3)$ of 3-regular graphs is an example of a class decidable
by a query algorithm that is definable in first-order logic but not by a
Boolean combination of universal sentences. Indeed, using the following facts
we get a query algorithm deciding whether a graph $G$ belongs to $\cls C(3)$
(by the way, with essentially the same proofs one gets the same results for
the class of $k$-regular graphs fo any $k\ge 2$).
\begin{itemize}
\item We check whether $G\in \cls C(\le 3)$, i.e., whether each vertex has
    at most 3 neighbors. Note that $\cls C(\le 3)$ is definable by a
    universal sentence. Hence there is a hom algorithm for $\cls C(\le 3)$,
    say consisting of $k\ge 1$, $F_1,\ldots, F_k$, and $X_{\le 3}\
    (\subseteq \mathbb  N^k)$.

\item We query $\hom(K_1, G)$ in order to get $n:= |V(G)|$.

\item We query $\hom(P_2, G)$, i.e., the number of homomorphisms from the
	path $P_2$ of two vertices to $G$. Then, $G$ is $3$-regular if and only
	if $\hom(P_2, G)= 3\cdot n$.
\end{itemize}
Hence, we have a hom algorithm for $\cls C(3)$ consisting of $k+2$,
$F_1,\ldots, F_k,K_1, P_2$, and $X$ where~$ X$ consists of all tuples
$(n_1,\ldots, n_k, n, 3n)$ with $(n_1,\ldots, n_k)\in X_{\le 3}$ and $n\ge
1$.

The class $\cls C(3)$ is definable in \FO\ by
\begin{align*}
\forall x\forall y_1\ldots \forall y_4\exists z_1\ldots \exists z_3
 \Big(\neg\big(\bigwedge_{1\le i< j\le 4} y_i\ne y_j \wedge \bigwedge_{i\in[4]}Exy_i\big)
 \wedge \big(\bigwedge_{1\le i< j\le 3} z_i\ne z_j \wedge
 \bigwedge_{j\in[3]}Exz_j\big)\Big).
\end{align*}
If $\cls C(3)$ would be definable by a Boolean combination of universal
sentences, then it would be definable by a sentence $\varphi$ of the form
\[
\varphi=\exists x_1\ldots \exists x_m\forall y_1\ldots \forall y_\ell\psi
\]
with $m, \ell\in \mathbb N$ and with quantifier-free $\psi$. Let $G$ be a
graph with more than $m+ 1$ vertices that is the disjoint union of copies of
the clique $K_4$. Of course, $G$ is $3$-regular. Hence, $G$ is a model of
$\varphi$. In particular, there are vertices $u_1, \ldots, u_m$ that satisfy
in $G$ the formula $\forall y_1\ldots \forall y_\ell \psi(x_1, \ldots, x_m)$
if we interpret $x_1$ by $u_1$, \ldots, $x_m$ by $u_m$. Choose a vertex $u\in
V(G)\setminus \{u_1, \ldots, u_m\}$. Then, $G\setminus u$, the graph induced
by~$G$ on $V(G)\setminus \{u\}$, is still a model of $\varphi$ but not
$3$-regular.
\end{rem}
By the previous remark the question arises whether every class $\cls
C(\varphi)$ for an \FO-sentence $\varphi$ of the form$\forall x_1\ldots
\forall x_m\exists y_1\ldots \exists y_\ell\psi$ with quantifier free $\psi$
can be decided by a query algorithm. We will see that already for the ``simplest''
formula  of this type, namely $\forall x\exists y Exy$,  the class $\cls C(\forall
x\exists y Exy)$, i.e., the class of graphs not containing isolated vertices,
has no query algorithm.

\section{Planarity and 3-colorability}\label{sec:plan}

As just mentioned we want to show that no query algorithm detects the
existence of an isolated vertex. In this section we prove the corresponding
result for planarity and 3-colorability, where some easy tools and
construction methods relevant in the much more involved proof for isolated
vertices are used. Note that the class of planar graphs and the class of
3-colorable graphs are definable in monadic second-order logic but not in
first-order logic.

By the following lemma a class has no query algorithm if there is no
emb algorithm for this class that only uses \emph{connected graphs}:
\begin{lem}\label{lem:congen}
Let $\cls C$ be a class of graphs. Assume that for every finite set $\cls K'$
of connected graphs there are graphs $G$ and $H$ such that (a) and (b) hold.
\begin{itemize}
\item[(a)] $G\in \cls C$ and $H\notin\cls C$.
		
\item[(b)] For all $F'\in \cls K'$ we have $\emb(F', G)= \emb(F', H)$.
\end{itemize}
Then there is no hom algorithm for $\cls C$, i.e., for every finite set $\cls
K$ of graphs there are graphs $G$ and~$H$ such that (c) and (d) hold
\begin{itemize}
\item[(c)] $G\in \cls C$ and $H\notin\cls C$.
		
\item[(d)] For all $F\in \cls K$ we have $\hom(F, G)= \hom(F, H)$.
\end{itemize}
\end{lem}

\proof By \eqref{eq:unionhom}, $\hom(F_1 \dotcup F_2, F_3)= \hom(F_1,
F_3)\cdot \hom(F_2, F_3)$ holds for arbitrary graphs $F_1, F_2, F_3$. Thus,
if for some $G$ and $H$ and for some class $\cls K$ we have $\hom(F, G)=
\hom(F, H)$ for all connected components of graphs in $\cls K$, then (d)
holds (for $\cls K$). Hence, we can assume that the graphs in $\cls K$ are
connected. Let $n:=\max\{|V(F)|$ $\mid F\in \cls K\}$ and
\[
\cls K':= \{F' \mid |V(F')|\le n \text{ \ and \  $F'$ is connected}\}
\]
(to ensure that $K'$ is finite, we just take exactly one copy of each
isomorphism type of such an $F'$). By assumption we know that there are
graphs $G$ and $H$ such that (a) and (b) hold for $\cls K'$. Now we
recall~\eqref{eq:homepiemb}, i.e.,
\[
\hom(F,G)= \sum_{F'\le F}\frac{1}{\aut(F')}\cdot\epi(F,F')\cdot \emb(F',G).
\]
If $F$ is connected, then $\epi(F,F') >0$ implies that $F'$ is connected,
too. That is, the values $\hom(F,G)$ for $F\in \cls K$ are determined by the
values of $\emb(F',G)$ for $F'\in \cls K'$. Therefore,~(d) holds by
(b).\proofend

\begin{theo}\label{thm:plan}
The class $\cls P$ of planar graphs and for $\ell\ge 2$ the class $\cls C(\ell\text{-col})$ of
$\ell$-colorable graphs have no query algorithm.
\end{theo}

\proof For a contradiction, by Lemma~\ref{lem:congen} we can assume that there is an
emb algorithm for~$\cls P$ that uses the finite set $\cls F$ of connected
graphs. By induction on $n:=|\cls F|$, we show that this cannot be the case.
Let $|\cls F|=1$, i.e., $\cls F= \{F\}$ for some connected graph $F$. We set
$ k:= |V(F)|+ 4$. Clearly, $K_k\notin \cls P$ and both, $\emb(F, F)$ and
$\emb(F, K_k) $, are non-zero. If $F$ is planar, we set (recall that for a
graph $G$ and $p\in \mathbb N$ by $pG$ we denote the disjoint union of $p$
copies of $G$),
\begin{eqnarray*}
G:= \emb(F, K_k) F
 & \text{and} &
H:= \emb(F, F) K_k
\end{eqnarray*}
By~\eqref{eq:unionemb}, $\emb(F,G)=\emb(F, K_k)\cdot \emb(F, F) =\emb(F,H)$
and $G\in \cls P$, $H\notin \cls P$, a contradiction. If~$F$ is not planar,
we set $G:= K_1$ and take as $H$ a topological minor of $K_k$ in which every
edge in $K_k$ is subdivided into $1+|(V(F)|$ edges. Then $H\notin \cls P$ but
$\emb(F, G)= 0 = \emb(F, H) $, again a contradiction.

Now assume $|\cls F|\ge 2$. If $\cls F$ contains no planar graphs, we
essentially proceed as in the preceding case: We take $G:= K_1$ and let $H$
be a topological minor of $K_k$ in which every edge is subdivided into
$1+\max\{|(V(F)|\mid F\in \cls F \}$ edges. Finally, assume that $\cls F$
contains a planar graph. Choose a ``minimal'' (w.r.t.\ $\le$) planar graph
$F\in\cls F$ and set $\cls F':= \cls F\setminus \{F\}$. By minimality,
$\emb(F', F)= 0$ for all planar $F'\in \cls F'$. As there is no embedding
from a non-planar graph to a planar graph, we have $\emb(F', F)=0$ for all
$F'\in \cls F'$. By induction hypothesis, there are $G_0$ and $H_0$
satisfying the desired properties with respect to $\cls F'$. If $\emb(F,
G_0)= \emb(F, H_0)$, then we can simply take $G:= G_0$ and $H:= H_0$.
Otherwise, assume first that $\emb(F, G_0)< \emb(F, H_0)$. We set
\[
G:=\aut(F)G_0 \dotcup (\emb(F, H_0)- \emb(F, G_0))F\quad
 \text{ and\ }\quad H:=\aut(F)H_0.
\]
Hence, $G\in \cls P$, $H\notin \cls P$, and $\emb(F,G)= \emb(F, H)$
(by~\eqref{eq:unionemb}). In case $\emb(F, G_0)> \emb(F, H_0)$ we argue
similarly.

We come to $\cls C(\ell\text{-col})$ with $\ell\ge 2$. Again for a
contradiction, we assume that there is an emb algorithm for~$\cls
C(\ell\text{-col})$ that uses a finite set $\cls F$ of connected graphs (by
Lemma~\ref{lem:congen}) and proceed by induction on $|\cls F|$. We consider
the case $|\cls F|=1$ and leave the induction step, which essentially
proceeds in the same way as for the class $\cls P$, to the reader.

Assume that $\cls F= \{F\}$ with a connected graph $F$. If $F$ is
$ \ell$-colorable, we set $k:= \ell+|V(F)|$, then both, $\emb(F, F)$ and $\emb(F,
K_k)$, are non-zero. We set
\begin{eqnarray*}
G:=\emb (F,K_k)F
 & \text{and} &
H:=\emb(F,F)K_k.
\end{eqnarray*}
Then $G\in \cls C(\ell\text{-col})$, $H\notin \cls C(\ell\text{-col})$ but
$\emb(G,F)= \emb(H,F)$.

Now assume that $F$ is not $\ell$-colorable, hence not 2-colorable, i.e, not
bipartite. Hence, $F$ contains a cycle of odd length, say of length $m$.
Choose a graph $H$ of girth (the length of a shortest cycle in $H$) greater
than $m$ and chromatic number (the least $s$ such that $H$ is $s$-colorable)
greater than~$\ell$. The existence of graphs of arbitrarily large girth and
chromatic number is due to Erd\H{o}s. By the first property $\emb(F,H)=0$ and
$ H\notin \cls C(\ell\text{-col})$ by the second property. With $G:= K_1$ we
have a graph in $\cls C(\ell\text{-col})$ with $\emb(F,G)=0$. \proofend

\section{No query algorithm detects isolated vertices}\label{sec:isol}

\begin{theo}\label{thm:isol}
The class $\cls C(\forall x\exists y Exy)$ of graphs without
isolated vertices has no query algorithm.
\end{theo}
In the proof we use some tools already
used in the preceding section:
\begin{itemize}
\item By Lemma~\ref{lem:congen} it suffices to show that $\cls
C(\forall x\exists y Exy)$ has no emb algorithm that only uses connected graphs.

\item If $\cls C$ is one of the classes $\cls P$, $\cls C(\ell\text{-col})$, or $\cls 
    x\exists y Exy)$, then the disjoint union $\dotbigcup_{i\in I}G_i$ of a
    family $(G_i)_{i\in I}$ of graphs is in $\cls C$ if and only if each
    $G_i$ is in $\cls C$.

\item By~(2), for graphs $F_1$ and $F_2$, $p, q\ge 1$, and a connected
    graph $G$,  we have
    \[
    \emb(G, pF_1\dotcup qF_2)=p\,\emb(G,F_1)+ q\,\emb(G,F_2),
    \]
    i.e., $\emb(G, pF_1\dotcup qF_2)$ is a linear combination of
    $\emb(G,F_1)$ and $\emb(G,F_2)$.
\end{itemize}
Furthermore note:
\begin{itemize}
\item Let $G$ be a connected graph with more than one vertex. If the graph
    $F'$ is obtained from the graph $F$ by adding a set of isolated
    vertices, then $\emb(G,F')= \emb(G,F)$.
\end{itemize}
\smallskip
\noindent{\em In this section (and only in this section) we fix an
enumeration of \emph{connected} graphs
\[
F_1, F_2, \ldots
\]
that contains exactly one copy of every isomorphism type of connected graphs
and respects the relation $\le$ (cf.~\eqref{eq:basord}), i.e., $F_i\le
F_{i+1}$ for all $i\ge 1$.}

\medskip
\noindent For $i\ge 1$ let $\alpha_i:= (\emb(F_i,F_j))_{j\ge 1}$ be the
vector containing the \emb-values of $F_i$ for connected graphs. We sketch
the idea underlying the proof of Theorem~\ref{thm:isol}. The central idea can
be vaguely expressed by saying that for each $n\ge 1$ there is an $r_n\in
\mathbb N$ such that appropriate sub-vectors of length $r_n$ of $r_n$-many of
the $\alpha_1,\ldots, \alpha_n$ are linearly independent vectors of the
vector space $\mathbb Q^{r_n}$ and hence, a basis of~$\mathbb Q^{r_n}$. In
particular, every further vector of~$\mathbb Q^{r_n}$ is a linear combination
of these vectors. Furthermore,~$r_n$ tends to infinity when $n$ increases.

For an arbitrary emb algorithm with connected graphs we must show the
existence of graphs $G$ and $H$, one with isolated vertices the other one
without, that cannot be distinguished by this emb algorithm. We use the tools
mentioned above to construct such graphs using the knowledge about the linear
independence or linear dependence of some tuples of vectors obtained in the
first steps of the proof.

\medskip

\noindent We turn to a proof of the result telling us that for each $n\ge
1$ appropriate finite sub-vectors of $\alpha_1, \ldots, \alpha_n$, whose
length tends to infinity when $n$ increases, are linearly independent.

\subsection*{Expressive graphs} We start with a definition.
\begin{defn}\label{def:expressive}
By induction on $s\ge 1$, we define whether $F_s$ is \emph{expressive}.
\begin{itemize}
\item $F_1$ is expressive (note that $F_1= K_1$).
		
\item Let $s\ge 2$. We set
	\begin{equation}\label{eq:Is-1}
	I_{s-1}:= \big\{i\bigmid \text{$2\le i\le s-1$ and $F_i$ is expressive}\big\}.
	\end{equation}
	Then $F_s$ is expressive if the matrix
	\[
	\big(\emb(F_i, F_j)\big)_{i\in \{1\}\,\cup\, I_{s-1},\, j\in I_{s-1}\cup\, \{s\}}
	\]
	is of full rank.
\end{itemize}
\end{defn}

\noindent For example, as $I_1= \emptyset$, $F_2\ (=P_2)$ (where $P_n$ denotes a path with $n$
vertices), and $\big(\emb(F_1, F_2)\big)= (2)$, we see that $F_2$ is
expressive. The relevant matrices for $F_3\ (=P_3)$ and $F_4\ (=K_3)$ are
\begin{eqnarray*}
\begin{pmatrix}
2 & 3 \\
2 & 4
\end{pmatrix}
& \text{and} &
\begin{pmatrix}
2 & 3 & 3 \\
2 & 4 & 6 \\
0 & 2 & 6
\end{pmatrix}.
\end{eqnarray*}
As the determinant of the latter matrix is zero, $F_4$ is not expressive.
%

\begin{lem}\label{lem:emblinearcomibination2}
Assume $F_s$ is not expressive. Then there are
\[
\big(p_j\big)_{j\in I_{s-1}\cup \{s\}}\in \mathbb Z^{|I_{s-1}\cup \{s\}|}
\]
with $p_s\ne 0$ such that for every $i\in \{1\}\cup I_{s-1}$,
\begin{equation*}
\sum_{j\in I_{s-1}\cup \{s\}} p_j \cdot \emb(F_i, F_j) = 0.
\end{equation*}
\end{lem}

\proof As $F_s$ is not expressive, we have $s\ge 4$. Let $t:=\max\, I_{s-1}$.
Then $I_{s-1}=I_{t-1}\cup\{t\}$. So $F_t$ is expressive and the matrix
\[
M_t:=\big(\emb(F_i, F_j)\big)_{i\in \{1\}\cup I_{t-1},\, j\in I_{t-1}\cup \{t\}}
\]
has maximal rank. Again as $F_s$ is not expressive, the matrix $M_s$, i.e.,
\[
\big(\emb(F_i, F_j)\big)_{i\in \{1\}\cup I_{s-1},\, j\in I_{s-1}\cup \{s\}} \ \
\Big(= \big(\emb(F_i, F_j)\big)_{i\in \{1\}\cup I_{t-1}\cup\{t\},\, j\in I_{t-1}\cup \{t,s\}}\Big)
\]
is not of full rank. Let $m:= \big|\{1\}\cup I_{s-1}\big|$. We consider the
columns of $M_s$ as vectors of the vector space $\mathbb Q^m$ over the
rationals. As the matrix $M_t$ has full rank, the first $m-1$ columns of
$M_t$ are linearly independent. This yields the claim of the lemma. \proofend


\begin{lem}\label{lem:expinfinite}
There are infinitely many expressive graphs.
\end{lem}

\proof Towards a contradiction, let $F_t$ be the expressive graph with
maximum index. Hence:
\begin{enumerate}
\item[(S1)] The matrix
    \[
    \big(\emb(F_i, F_j)\big)_{i\in \{1\}\cup I_{t-1},\, j\in I_{t-1}\cup \{t\}}
    \]
    is of full rank.
	
\item[(S2)] For any $s> t$
	\[
	\big(\emb(F_i, F_j)\big)_{i\in \{1,t\}\cup I_{t-1},\, j\in I_{t-1}\cup \{t,s\}}
	\]
    is not of full rank.
\end{enumerate}
For every $i\ge 1$, let $\alpha_i$ be the following infinite (row) vector
\[
\alpha_i:= \big(\emb(F_i, F_j)\big)_{j\ge 1}.
\]
For a non-empty finite set $J$ of positive
natural numbers let
\[
\subvec{\alpha_i}{J}:=(\emb(F_i, F_j))_{j\in J}.
\]
denote the sub-vector of $\alpha_i$ obtained by restricting to the
coordinates with index in $J$.
We view $\subvec{\alpha_i}{J}$ as an (row) vector of the vector space $\mathbb Q^{|J|}$.
 Set
\begin{eqnarray*}
I^t:= \{1\}\cup I_{t-1} & \text{and} & J^t:= I_{t-1}\cup \{t\}.
\end{eqnarray*}
Furthermore, assume
\[
\text{$I^t= \{i_1, \ldots, i_r\}$ with $i_1<\ldots <i_r$}.
\]
Then,
\[
\big(\emb(F_i, F_j)\big)_{i\in I^t,\, j\in J^t}=
\begin{pmatrix}
	\subvec{\alpha_{i_1}}{J^t} \\
	\vdots \\
	\subvec{\alpha_{i_r}}{J^t}
\end{pmatrix}.
\]
Therefore by (S1),
\begin{equation}\label{eq:claim1}
\text{$\subvec{\alpha_{i_1}}{J^t}, \ldots,
\subvec{\alpha_{i_r}}{J^t}$ are linearly independent vectors in $\mathbb Q^r$ }
\end{equation}
and hence, a basis of~$\mathbb Q^r$. Now consider $\subvec{\alpha_t}{J^t}=
\big(\emb(F_t, F_j)\big)_{j\in J^t}$, also a vector in $\mathbb Q^r$. Thus,
\begin{equation}\label{eq:ci}
\text{there are \emph{unique} $c_1, \ldots,
    c_r\in \mathbb Q$  such that} \ \ \subvec{\alpha_t}{J^t}
 = \sum_{\ell\in [r]} c_{\ell}\cdot \subvec{\alpha_{i_{\ell}}}{J^t}.
\end{equation}

\medskip \noindent \textit{Claim 1.} For every $s> t$,
\[
\subvec{\alpha_t}{J^t\cup \{s\}}
= \sum_{\ell\in [r]} c_{\ell}\cdot \subvec{\alpha_{i_{\ell}}}{J^t\cup \{s\}}.
\]
In particular,
\[
\emb(F_t, F_{s})= \sum_{\ell\in [r]} c_{\ell}\cdot \emb(F_{i_{\ell}}, F_{s}).
\]

\medskip
\noindent \textit{Proof of Claim 1:} The $(r+1)\times (r+1)$ matrix in (S2)
is precisely
\[
\begin{pmatrix}
	\subvec{\alpha_{i_1}}{J^t\cup \{s\}} \\
	\vdots \\
	\subvec{\alpha_{i_r}}{J^t\cup \{s\}} \\[2mm]
	\subvec{\alpha_{t}}{J^t\cup \{s\}}
\end{pmatrix}.
\]
By (S2), it has rank $\le r$. By~\eqref{eq:claim1},
\[
\subvec{\alpha_{i_1}}{J^t\cup \{s\}}, \ldots,
\subvec{\alpha_{i_r}}{J^t\cup \{s\}}
\]
are linearly independent as well. Therefore, for some $c'_1, \ldots, c'_r\in
\mathbb Q$ we have
\begin{equation}\label{eq:cpi}
	\subvec{\alpha_t}{J^t\cup \{s\}}
	= \sum_{\ell\in [r]} c'_{\ell}\cdot \subvec{\alpha_{i_{\ell}}}{J^t\cup \{s\}}.
\end{equation}
This further implies
\begin{equation*}
	\subvec{\alpha_t}{J^t}
	= \sum_{\ell\in [r]} c'_{\ell}\cdot \subvec{\alpha_{i_{\ell}}}{J^t}.
\end{equation*}
By the uniqueness claim in~\eqref{eq:ci} we conclude
$
c'_{\ell}= c_{\ell}
$
for all $\ell\in [r]$. With~\eqref{eq:cpi} we have shown the claim.
\hfill$\dashv$

\bigskip \noindent \textit{Claim 2.} $c_1\ne 0$.

\medskip
\noindent \textit{Proof of Claim 2:} Otherwise, by Claim~1, the rank of the $r\times (r+1)$ matrix
\[
\begin{pmatrix}
	\subvec{\alpha_{i_2}}{J^t\cup \{s\}} \\
	\vdots \\
	\subvec{\alpha_{i_r}}{J^t\cup \{s\}} \\[2mm]
	\subvec{\alpha_{t}}{J^t\cup \{s\}}
\end{pmatrix}
\]
is at most $r-1$. However, this matrix contains as a submatrix
\[
\big(\emb(F_i, F_j)\big)_{i,j\in J^t},
\]
which has rank $r$ as it is upper triangular with positive elements in the
diagonal. \hfill~$\dashv$


\medskip

\noindent
By Claim~1 and Claim~2, there are
\[
\big(c'_i\big)_{i\in \{1,t\}\cup I_{t-1}}
 \in \mathbb Q^{|\{1,t\}\cup I_{t-1}|}
\]
with $c'_1\ne 0$ such that for all $s>t$ \big(by $\{1,t\}\cup I_{t-1}= \{i_1, \ldots, i_r
\}\cup \{t\}$\big)
\[
\sum_{i\in \{1,t\}\cup I_{t-1}} c'_i \cdot \emb(F_i, F_{s})
 =\sum_{ i\in \{i_1,\ldots, i_r \}\cup\{t \}} c'_i \cdot \emb(F_i, F_{s})= 0.
\]

\medskip
\noindent Finally we show that this equality cannot hold if $F_s$ is a
sufficiently large clique. Assume $F_{s}= K_m$ for some $m\ge 1$ to be
determined later. Then for every $i\ge 1$
\[
\emb(F_i, F_{s})= |V(F_i)|!\cdot \binom{m}{|V(F_i)|}.
\]
Therefore,
\begin{align*}
0 & = \sum_{i\in \{1,t\}\cup I_{t-1}} c'_i \cdot \emb(F_i, F_{s})
 = \sum_{n\ge 1}\ \sum_{\substack{i\in \{1,t\}\cup I_{t-1}\\
  |V(F_i)|= n}} c'_i \cdot n!\cdot \binom{m}{n} \\[1mm]
 & = \sum_{n\ge 1} n!\cdot \binom{m}{n}
  \cdot \sum_{\substack{i\in \{1,t\}\cup I_{t-1}\\ |V(F_i)|= n}} c'_i
 \  = \ \sum_{n\ge 1} n!\cdot \binom{m}{n}\cdot e_n,
\end{align*}
where
\[
e_n:= \sum_{\substack{i\in \{1,t\}\cup I_{t-1}\\ |V(F_i)|= n}} c'_i.
\]
Note that $e_1= c'_1\ne 0$. Hence there exists a maximum $n\ge 1$ with
$e_n\ne 0$. Clearly,  $n\le |V(F_t)|$. Thus,
\[
n!\cdot \binom{m}{n}\cdot e_n
= -\sum_{n> \ell\ge 1} \ell!\cdot \binom{m}{\ell}\cdot e_{\ell}
\]
As $e_n\ne 0$, we conclude
$
\Theta(m^n)= O(m^{n-1}),
$
which cannot hold for sufficiently large $m$, a contradiction. \proofend

\subsection*{The class $\cls C(\forall x\exists y Exy)$ of graphs without isolated vertices}\smallskip

\noindent We turn to the proof of Theorem~\ref{thm:isol}. By Lemma~\ref{lem:congen} we
know that it suffices to show:
\begin{lem}
For every finite set $\cls K$ of connected graphs there are graphs $G$ and
$H$ such that (a) and (b) hold.
\begin{itemize}
\item[(a)] $G$ has an isolated vertex and $H$ does not.
		
\item[(b)] For all $F\in\cls K$ we have $\emb(F, G)= \emb(F, H)$.
\end{itemize}
\end{lem}

\proof Clearly, we can assume that $\cls K$ only contains graphs of the
enumeration $F_1, F_2, \ldots$ of connected graphs. We prove the claim by
induction on the number $k$ of graphs in $\cls K$. First assume that $k=1$,
i.e., $\cls K= \{F\}$ for some graph $F$. If $F= K_1$, then we can take $G:=
K_1\dotcup K_1$ and $H:= P_2$. Otherwise, $F$ contains at least one edge
since it is connected. Then let $G:= K_1\dotcup P_2$ and $H:= P_2$.

\medskip
\noindent Now let $k\ge 2$. We distinguish two cases.
\paragraph{Case 1: All graphs in $\cls K$ are expressive.} By
Lemma~\ref{lem:expinfinite} we can find an $s$ such that $F_s$ is expressive
and $I_{s-1}$ \big(as defined in~\eqref{eq:Is-1}\big) contains all the
indices of the graphs in $\cls K$.

As the matrix $\big(\emb(F_i,F_j)\big)_{i,j\in I_{s-1}}$ is upper triangular
with positive diagonal elements, it has maximal rank. Hence,  for some $(p_j)_{j\in
I_{s-1}\cup \{s\}}\in \mathbb Z^{|I_{s-1}|+1}$ with $p_s\ne 0$ and all $i\in
I_{s-1}$,
\begin{equation}\label{eq:Scolumns}
\sum_{j\in I_{s-1}\cup \{s\}} p_j \cdot \emb(F_i, F_j) = 0.
\end{equation}
As $F_s$ is expressive, by Definition~\ref{def:expressive} the matrix
$
\big(\emb(F_i, F_j)\big)_{i\in \{1\}\cup I_{s-1},\, j\in I_{s-1}\cup \{s\}}
$
is of full rank, or equivalently, its column vectors are linearly
independent. Combined with~\eqref{eq:Scolumns} we get (recall $1\notin
I_{s-1}$)
\[
\sum_{j\in I_{s-1}\cup \{s\}} p_j \cdot \emb(F_1, F_j)\ne 0.
\]
As $F_1= K_1$, this is equivalent to
\begin{equation}\label{eq:sizeGH}
\sum_{\substack{j\in I_{s-1}\cup \{s\}\\ p_j\ge 0}} p_j \cdot |V(F_j)|
 \ne \sum_{\substack{j\in I_{s-1}\cup \{s\}\\ p_j< 0}} -p_j \cdot |V(F_j)|.
\end{equation}
We set
\begin{eqnarray*}
G_0:= \underset{{\substack{j\in I_{s-1}\cup \{s\}\\ p_j\ge 0}}}{\dotbigcup}\
 p_jF_j,
 & &
H_0:= \underset{{\substack{j\in I_{s-1}\cup \{s\}\\ p_j< 0}}}{\dotbigcup}\
(-p_j)F_j.
\end{eqnarray*}
Since $1\notin I_{s-1}\cup \{s\}$, neither $G_0$ nor $H_0$ contains an
isolated vertex. Furthermore, by~\eqref{eq:sizeGH}
$
|V(G_0)|\ne |V(H_0)|$.
Without of loss of generality, assume $d:= |V(H_0)|- |V(G_0)|> 0$. Set
\begin{eqnarray*}
G:= G_0\dotcup dK_1
 & \text{and} &
H:= H_0.
\end{eqnarray*}
Then $G$ and $H$ have the same number of vertices. Clearly $G$ contains
isolated vertices, while $H$ does not; hence, (a) holds.

For~(b) let $F\in\cls K$, say $F=F_i$. As $s> i$ and $F_{i}$ is expressive,
either $i= 1$ or $i\in I_{s-1}$. For $i=1$,
\[
\emb(F_1, G)= |V(G)|= |V(H)|= \emb(F_1, H).
\]
For $i\in I_{s-1}$,
\begin{align*}
\emb(F_{i}, G) &= \emb(F_{i}, G_0) & \text{\big(as $F_{i}$ is connected\big)} \\
 & = \sum_{\substack{j\in I_{s-1}\cup \{s\}\\ p_j\ge 0}} p_j\cdot \emb(F_{i}, F_j)
  & \text{\big(again as $F_{i}$ is connected and \eqref{eq:unionemb}\big)} \\
 & = \sum_{\substack{j\in I_{s-1}\cup \{s\}\\ p_j< 0}} -p_j\cdot \emb(F_{i}, F_j)
  & \text{\big(by~\eqref{eq:Scolumns}\big)} \\
 & = \emb(F_{i}, H_0)= \emb(F_{i}, H).
\end{align*}
Thus, (b) holds too.

\paragraph{Case 2: Some $F\in \cls K$ is not expressive.}
Let $s$ be the \emph{minimum} integer such that $F_{s}\in \cls K$ and~$F_s$
is not expressive. Recall that we prove our claim by induction on the number
$k$ of graphs in $\cls K$. Hence there are
\[
i_1,i_2,\ldots, i_{t-1}, i_{t+1}, \ldots, i_k
\ \ \text{ with\ }
i_1<i_2< \cdots < i_{t-1}< s < i_{t+1}< \cdots < i_k
\]
such that $F_{i_1}, \ldots, F_{i_{t-1}}, F_s, F_{i_{t+1}}, \ldots, F_{i_k}$
are the graphs in $\cls K$. By induction hypothesis, there are two graphs
$G_0$ and $H_0$ such that
\begin{itemize}
\item[(E1)] $G_0$ has isolated vertices and $H_0$ does not.
	
\item[(E2)] For all $r\in [k]\setminus \{t\}$, \ \
	$
	\emb(F_{i_{r}}, G_0)= \emb(F_{i_{r}}, H_0)$.
\end{itemize}
If $\emb(F_{s}, G_0)= \emb(F_{s}, H_0)$, then we are already done. Otherwise,
\begin{itemize}
\item[(E3)] $g:= \emb(F_{s}, G_0) \ne \emb(F_{s}, H_0) =:h$.
\end{itemize}
In addition, we observe that
\begin{itemize}
\item[(E4)] $F_{i_1}, \ldots, F_{i_{t-1}}$ are all expressive.
\end{itemize}
Since $F_s$ is not expressive, by
Lemma~\ref{lem:emblinearcomibination2} there exist
$
\big(p_j\big)_{j\in I_{s-1}\cup \{s\}}\in \mathbb Z^{|I_{s-1}\cup \{s\}|}
$
with $p_s\ne 0$ such that for every $i\in \{1\}\cup I_{s-1}$,
\begin{equation}\label{eq:matrixnoexpressive}
\sum_{j\in I_{s-1}\cup \{s\}} p_j \cdot \emb(F_i, F_j) = 0.
\end{equation}
We set
\[
G_1:= \underset{{\substack{j\in I_{s-1}\cup \{s\}\\ p_j\ge 0}}}{\dotbigcup}\
p_jF_j\quad \text{ and}\quad H_1:= \underset{{\substack{j\in I_{s-1}\cup \{s\}\\ p_j< 0}}}{\dotbigcup}\
(-p_j)F_j.
\]
Without loss of generality, we assume $p_s> 0$. Hence $G_1$ contains $p_s$
disjoint copies of $F_s$ while $H_1$ contains none.

\bigskip \noindent
\textit{Claim 1.} Neither $G_1$ nor $H_1$ contains an isolated vertex.

\medskip
\noindent \textit{Proof of Claim 1:} For each $j\in I_{s-1}\cup \{s\}$ the
graph $F_j$ is connected and contains at least two vertices. \hfill$\dashv$

\bigskip \noindent
\textit{Claim 2.} For every $r\in [k]\setminus \{t\}$, we have
$\emb(F_{i_{r}}, G_1)= \emb(F_{i_{r}}, H_1)$.

\medskip
\noindent \textit{Proof of Claim 2:} First, consider the case $r< t$.
Hence, $i_{r}< s$. Then (E4) implies that $i_{r}\in \{1\}\cup I_{s-1}$. It
follows that
\begin{align*}
\emb(F_{i_{r}}, G_1)
 & = \sum_{\substack{j\in I_{s-1}\cup \{s\}\\ p_j\ge 0}} p_j\cdot \emb(F_{i_{r}}, F_j)
  & \text{\big(by \eqref{eq:unionemb}\big)} \\
 & = \sum_{\substack{j\in I_{s-1}\cup \{s\}\\ p_j< 0}} -p_j\cdot \emb(F_{i_{r}}, F_j)
  & \text{\big(by~\eqref{eq:matrixnoexpressive}\big)} \\
 & = \emb(F_{i_{r}}, H_1).
\end{align*}
Now let $t< r\le k$. Then $s< i_r$ and hence, $\emb(F_{i_{r}}, F_j)= 0$ for
$j\in[s]$. Therefore,
\[
\emb(F_{i_{r}}, G_1)= \emb(F_{i_{r}}, H_1).
\]
This proves our claim. \hfill$\dashv$

\bigskip
\noindent \textit{Claim 3.} $\emb(F_s, G_1)= p_s\cdot \aut(F_s)> 0$ and
$\emb(F_{s}, H_1)= 0$.

\medskip
\noindent \textit{Proof of Claim 3:} Note that $j< s$ for $j\in I_{s-1}$.
Thus
\begin{align*}
\emb(F_s, G_1)
 = \sum_{\substack{j\in I_{s-1}\cup \{s\}\\ p_j\ge 0}} p_j\cdot \emb(F_s, F_j)
 = p_s\cdot \emb(F_s, F_s)
 = p_s\cdot \aut(F_s).
\end{align*}
Arguing similarly we get $\emb(F_s, H_1)= 0$. \hfill$\dashv$

\medskip
\noindent Using the graphs $G_0$ and $H_0$ satisfying (E1), finally we define
the graphs $G$ and $H$. Thereby, for~$g$ and $h$ defined in (E3) we first
assume $g> h$. We set
\[
G := \emb(F_{s},
		G_1)G_0  \dotcup (g-h)H_1 \quad \text{ and} \quad H := \emb(F_{s}, G_1)H_0 \dotcup (g-h)G_1.
\]
By (E1),  $G$ contains  isolated vertices while $H$ does
not by Claim~1. This proves (a). To establish~(b), let $r\in
[k]\setminus \{t\}$. Then
$
\emb(F_{i_{r}}, G)= \emb(F_{i_{r}}, H)
$
follows from (E2) and Claim~2. For $F_s$ we get
\begin{align*}
\emb(F_{s}, G)
 & = \emb(F_{s}, G_1)\cdot \emb(F_{s}, G_0)+ (g- h)\cdot \emb(F_{s}, H_1)
  & \text{\big(by the definition of $G$\big)} \\
 & = g\cdot \emb(F_{s}, G_1) + 0
  & \text{\big(by (E3) and Claim~3\big)} \\
 & = h\cdot \emb(F_{s}, G_1) + (g-h) \cdot \emb(F_{s}, G_1) \\
 & = \emb(F_{s}, G_1)\cdot \emb(F_{s}, H_0)+
  (g- h)\cdot \emb(F_{s}, G_1)
  & \text{\big(by (E3)\big)} \\
 & = \emb(F_{s}, H) & \text{\big(by the definition of $H$\big)}.
\end{align*}
%
If $g< h$, we
 argue similarly. \proofend

\section{On the way to adaptive homomorphism counts}\label{sec:star}

By the \LV\ Isomorphism Theorem, for a graph $G$ the values $\hom(F,G)$ for
the graphs~$ F$ with $F\le G$ determine~$G$ (up to isomorphism) and thus we
know whether~$G$ has an isolated vertex. The next result shows that instead
of all these $F$'s it suffices to consider stars with at most as many
vertices as $G$ has. Let $S_j$ denote the star of~$j$ vertices, i.e., a graph
that consists of a vertex of degree $j-1$ (the center of the star) and $j-1$
vertices of degree~$1$, all neighbors of the center. For a vertex $u$ of a
graph we denote by $\deg(u)$ its degree. Note that $\deg(u)= 0$ means
that~$u$ is isolated.

\begin{prop}\label{pro:star1}
Let $G$ be a graph and $d_i:= \big|\{u\in V(G) \mid \deg(u)=
i\}\big|$ for $i\ge 0$. If $n:=|V(G)|$, then  the sequence of values of\/ $\hom(S_j, G)$ for
$j\in[n]$ determines $d_0, \ldots, d_{n-1}$ (note that $d_k=0$ for $k\ge n$).
\end{prop}

\proof We use the following well known observation: For $v\in V(G)$ the
number of homomorphisms sending  the center of the star $S_j$ to $v$ is
$\deg(v)^{j-1} $. Hence, for $j\in [n]$ we have
\begin{equation}\label{eq:starhom}
\hom(S_j, G)= \sum_{0\le i\le n-1} d_i\cdot i^{j-1}.
\end{equation}
We consider the equation \eqref{eq:starhom} for $j=2,\ldots,n$. They form a system of linear equations
in the unknowns $d_1,\ldots,d_{n-1}$. Its matrix is the Vandermonde matrix
\[
\begin{pmatrix}
1 & 2^1 & \ldots & (n-1)^1 \\[2mm]
1 & 2^2 & \ldots & (n-1)^2 \\[2mm]
& & \ddots & \\[2mm]
1 & 2^{(n-1)} & \ldots & (n-1)^{(n-1)}
\end{pmatrix}.
\]
As this matrix is invertible, the system of equations determines $d_1,\ldots,
d_{n-1}$ and therefore,~$d_0$.\proofend It turns out that for graphs of a
fixed number of vertices we get the numbers of vertices of the same degree by
a single homomorphism count:
\begin{prop}\label{pro:star2}
Fix $n\ge 2$.  Then for graphs $G$ with $|V(G)|=n$ the homomorphism count
$\hom(S_{n\cdot\log n}, G)$ determines the sequence $d_0, \ldots, d_{n-1}$ of
$G$.
\end{prop}
\proof
We order the tuples in $\mathbb N^n$ as follows: if $x,y\in \mathbb N^n$,
$x=(x_0, \ldots, x_{n-1})$ and $y= (y_0, \ldots, y_{n-1})$, then $x< y$ if
for some $i$ with $0\le i \le n-1$
\begin{eqnarray}\label{eq:lex}
\text{$x_\ell=y_\ell$ for $i+1\le \ell\le n-1$}
 & \text{and} & x_i< y_i.
\end{eqnarray}
If $\sum_{0\le j\le n-1} x_j= \sum_{0\le j\le n-1} y_j= n$ and $x< y$, then
the $i$ satisfying~\eqref{eq:lex} cannot be $0$ as then $x_0= n-
\sum_{j\in[n-1]} x_j= n- \sum_{j\in [n-1]}y_j = y_0$.\smallskip

We set $x(G):= (d_0,\ldots, d_{n-1})$ and define $x(H)$ for a graph~$H$ with
$n=|V(H)|$ analogously. It suffices to show that
\[
\text{$x(G)< x(H)$ \ \
  implies \ \
  $\hom(S_{n\cdot\log n}, G)< \hom(S_{n\cdot\log n}, H)$}.
\]
By \eqref{eq:starhom} the next claim yields the statement of the
proposition.

\medskip
\noindent \textit{Claim}. If $x(G)< y$ and $\sum_{0\le j\le n-1}y_j=n$, then
$\hom(S_{n\cdot \log n}, G)$ $<\sum_{0\le j\le n-1} y_j\cdot j^{n\cdot \log
n-1}.$

\medskip
\noindent \textit{Proof of the Claim}. As $x(G)< y$ and $\sum_{j\in[n]}y_j=n$, choose $i\in [n-1]$
such that \eqref{eq:lex} holds. Then,
\begin{align*}
 & \hom(S_{n\cdot \log n}, G) = \sum_{j\in[n-1]} d_j\cdot j^{n\cdot \log n -1}\\
  & \le \sum_{i+1\le j\le n-1}  y_j\cdot j^{n\cdot \log n -1}
   + ( y_i-1)\cdot i^{n\cdot \log n -1}
   + \sum_{1\le j\le i-1} d_j\cdot j^{n\cdot \log n -1}\\[1mm]
 & \le \sum_{i+1\le j\le n-1}  y_j\cdot j^{n\cdot \log n -1}
  +  y_i\cdot i^{n\cdot \log n -1}
  - i^{n\cdot \log n -1}
  + (n-1) (i-1)^ {n\cdot \log n -1}\\[1mm]
  & < \sum_{i+1\le j\le n-1} y_j\cdot j^{n\cdot \log n -1}
   + y_i\cdot i^{n\cdot \log n -1}\\
  & \hspace*{5.6cm}
   \text{\big(by $i^{n\cdot \log n-1}> (n-1)(i-1)^{n\cdot \log n-1}$; see below\big)}\\[1mm]
 & \le \sum_{j\in [n-1]} y_j\cdot j^{n\cdot \log n -1}.
\end{align*}
It remains to show for $0<i<n$,
\begin{equation}\label{eq:dlogn}
i^{n\cdot \log n-1}>(n-1)\cdot(i-1)^{n\cdot \log n-1}.
\end{equation}
Clearly this holds for $i=1$ and for $i=2$. In fact, for $i=2$ the right hand
side is equal to $n-1$. As then $n\ge 3$ and $\log 3> 1$ the left hand side
is greater than $2^{n-1}$. Clearly, for $n\ge 3$, we have $2^{n-1}> n-1$.
Hence, we assume $2< i< n$. The inequality~\eqref{eq:dlogn} is equivalent to
\[
1> (n-1)\cdot \Big(1-\frac{1}{i}\Big)^{n\cdot \log n-1}.
\]
We show that the right hand side is less than 1:
\begin{align*}
(&n-1)\cdot \Big(1-\frac{1}{i}\Big)^{n\cdot \log n-1}
= (n-1)\cdot \Big(\Big(1-\frac{1}{i}\Big)^i\Big)^{\frac{n\cdot \log n-1}{i}} \\[1mm]
& < (n-1)\cdot \left(e^{-1}\right)^{\frac{n\cdot \log n-1}{i}}
\qquad \text{(as $\Big(1-\frac{1}{i}\Big)^i<e^{-1}$)} \\[1mm]
& = \frac{(n-1)}{e^{\frac{n\cdot \log n-1}{i}}}
= \frac{(n-1)\cdot e^{\frac{1}{i}}}{e^{\frac{n\cdot \log n}{i}}}
= \frac{(n-1)\cdot e^{\frac{1}{i}}}{n^{\frac{n}{i}}} \\[1mm]
& < \frac{(n-1)\cdot e^{\frac{1}{i}}}{(n-1)^{\frac{n}{i}}}
= \frac{e^{\frac{1}{i}}}{(n-1)^{(n-i)\cdot\frac{1}{i}}}
= \Big( \frac{e}{(n-1)^{n-i}}\Big)^{\frac{1}{i}} \\[1mm]
& < 1 \qquad \qquad\text{ (as $n\ge 4$ as $2< i< n$; thus,
	 $\frac{e}{(n-1)^{n-i}}<1$)}.\benda
\end{align*}

\begin{rem}
For $k,d\in \mathbb N$ there is a query algorithm deciding whether there are
exactly $k$ elements of degree greater or equal to $d$ (this might surprise
in view of the result concerning isolated vertices presented in the previous
section). In fact, there is an existential FO-sentence $\varphi_{k,d}$
expressing ``there are at least $k$ elements of degree greater or equal to
$d$'' and thus the universal sentence $\neg\varphi_{k,d}$ expresses ``there
are less than $k$ elements of degree greater or equal to $d$.'' Hence
$\varphi_{k,d}\wedge \neg\varphi_{k+1,d}$ expresses ``there are exactly $k$
elements of degree greater or equal to $d$.''
\end{rem}

\section{Adaptive hom algorithms}\label{sec:ada}
In Section~\ref{sec:isol} we showed that there is no query algorithm that
decides whether a graph $G$ is in $\cls C(\forall x\exists y Exy)$, i.e.,
whether $d_0= 0$ for $G$. However, Proposition~\ref{pro:star2} shows that for
a graph $G$ with $n$ vertices this can be decided by querying $\hom(S_{n\cdot
\log n},G)$. Thus, we have an algorithm for $\cls C(\forall x\exists y Exy)$
consisting of two homomorphism counts:
\begin{itemize}
\item query $n:= \hom(K_1,G) \ \ (=|V(G)|)$,

\item query $\hom(S_{n\cdot \log n}, G)$.
\end{itemize}
That is, the selection of the graph for the second homomorphism count, in our
case $S_{n\cdot \log n}$, depends on the answer to the first query. This
leads to the notion of adaptive hom algorithm. By~\textsc{Graph} we denote
the class of all graphs.

\begin{defn}\label{def:ada}
Let $\cls C$ be a class of graphs and $k\ge 1$. A \emph{$k$ adaptive hom
algorithm for $\cls C$} consists of a  graph $F$ (the \emph{starting graph})
and of function $N: \bigcup_{i \in [k-1]} \mathbb N^i \to \textsc{Graph}$
(the \emph{next graph function}) and a subset $X$ of $\mathbb N^k$ such that
for all~$G$,
\[
G\in \cls C\iff (n_1,\ldots, n_k)\in X,
\]
where $n_1:= \hom(F, G)$ and
\[
n_2:= \hom(N(n_1), G),\ \ n_3:=\hom(N(n_1,n_2), G),\ \
\ldots, \ \ n_k:= \hom(N(n_1, n_2, \ldots, n_{k-1}), G).
\]
We then say that
\emph{$\cls C$ can be decided by a $k$ adaptive hom algorithm}.
\end{defn}

For example, by the previous remarks we have a 2 adaptive hom algorithm for
$\cls C(\forall x\exists y Exy)$ consisting of starting graph $K_1$, the next
graph function $N:\mathbb N\to \textsc{Graph}$ with $N(n):= S_{n\cdot \log
n}$, and the set $X\subseteq \mathbb N^2$ contains a tuple $(n,m)$ if and
only if
\[
\text{$n\ge 1$ \ and \ $m=\sum_{j\in [n-1]}d_j\cdot j^{n\cdot \log n-1}$
 \ with \ $\sum_{j\in [n-1]}d_j= n$}.
\]

\noindent The main result of this section reads as follows:
\begin{theo}\label{thm:3adapsuf}
Every class of graphs can be decided by a $3$ adaptive hom algorithm.
\end{theo}

\noindent To get this result it suffices to show that for fixed $n$ there are
two graphs whose homomorphism counts determine the isomorphism type of graphs
with $n$ vertices, more precisely:
\begin{theo}\label{thm:3adapiso}
Let $n\ge 1$. Then there exist graphs $F_1\ (=F_1(n))$ and $F_2\ (=F_2(n))$
such that for all graphs $G$ and $H$ with $|V(G)|=|V(H)|=n$,
\begin{center}
$\hom(F_i,G)= \hom(F_i,H)$ \ for $i\in[2]$ \ \ imply \ \ $G\cong H$.
\end{center}
\end{theo}

\noindent In fact, if we assume this result, then for an arbitrary class
$\cls C$ of graphs we get a 3 adaptive hom algorithm taking $K_1$ as starting
graph $F$, the next graph function $N: \mathbb N\cup \mathbb N^2\to
\textsc{Graph}$ with $N(n)= F_1(n)$ and $N(n,n')= F_2(n)$, and
\[
X:= \{\big(|V(G)|, F_1(|V(G)|),F_2(|V(G)|)\big) \mid G\in \str{ C}\}.
\]

\begin{cor}\label{cor:3iso}
For all graphs $G$ and $H$,
\begin{align*}
\text{if $n_0:=\hom(K_1, G)= \hom(K_1, H)$}, &
\text{ $\hom(F_1(n_0), G)= \hom(F_1(n_0), H)$}, \\
 & \text{and $\hom(F_2(n_0), G)= \hom(F_2(n_0), H)$, then $G\cong H$}.
\end{align*}
\end{cor}

\noindent Hence, by ``three adaptive hom-queries'' we can characterize the
isomorphism type of any graph. In Corollary~\ref{cor:twonot} we will see that
it is not possible to do this by two queries in general.

\medskip
\noindent We turn to a proof of Theorem~\ref{thm:3adapiso}. An important tool
in the proof will be the following lemma.

\begin{lem}\label{lem:encoding}
Let $n\ge 1$ and $\cls K$ be a finite set of graphs. We can construct a graph
$F_{\cls K}$ such that for all graphs $G$ and $H$ with exactly $n$ vertices
we have $\hom(F_{\cls K},G) = \hom(F_{\cls K}, H)$ if and only if $G$ and $H$
satisfy at least one of the conditions (a) and (b).
\begin{enumerate}
\item[(a)] There exist $F, F'\in\cls K$ such that
\begin{eqnarray*}
\hom(F, G)=0 & \text{and} & \hom(F', H)= 0.
\end{eqnarray*}
		
\item[(b)] For all $F\in \cls K$,
\[
\hom(F, G) = \hom(F, H).
\]
\end{enumerate}
\end{lem}

\proof The idea of the construction is best seen by assuming $\cls K=\{F_1,
F_2\}$ (iterating the following process one gets the general case). Let
\[
r:= n^{|V(F_1)|}.
\]
We set $F_{\cls K}:= F_1\dotcup rF_2$. By~\eqref{eq:unionhom} for every graph
$F$,
\begin{equation}\label{eq:huge}
\hom(F_{\cls K}, F)= \hom(F_1, F)\cdot \hom(F_2, F)^r.
\end{equation}
Furthermore, $\hom(F_1, F)\le r $if $|V(F)|\le n$.

Now let $G$ and $H$ be graphs with exactly $n$ vertices.
If~(a) or (b) hold, then $\hom(F_{\cls K},G) = \hom(F_{\cls K}, H)$.
Conversely, assume $z:=\hom(F_{\cls K},G) = \hom(F_{\cls K}, H)$. If $z=0$,
then (a) must hold by~\eqref{eq:huge}. If $z=1$, then $\hom(F_i, G) =
\hom(F_i, H) =1$ for $i\in[2]$ and (b) holds. Otherwise, $z\ge 2$. Let $F:=
G$ or $F:= H$ and set $x:= \hom(F_1, F)$ and $y:= \hom(F_2, F)$. Let $p$ be a
prime number with $p|z$ (i.e., $p$ divides $z$). Choose the maximum $k$ such
that $p^k|z$. Write $k$ in the form $k=\ell\cdot r+ m$ with $0\le m<r$. As
$z= x\cdot y^r$ and $x\le r$, the factor $p^m$ appears in $x$ and the factor
$p^\ell$ in~$y$. This determines $x$ and $y$; they do not depend on the $F\in
\{G,H \}$ chosen. \proofend

If the alternative (b) holds, we can replace in a hom algorithm the
set $\cls K$ of graphs by the single graph $F_{\cls K}$. The previous lemma
doesn't help too much if the alternative (a) holds. To handle this
alternative essentially we consider the bipartite graphs and the
non-bipartite graphs separately. For this purpose we recall the definition
and some simple facts on bipartite graphs.

A graph $G$ is \emph{bipartite} if there is a partition $V(G)=X\dotcup Y$
such that each edge of $G$ has one end in $X$ and one end in $Y$. The
following lemma contains some simple facts on bipartite graphs.
\begin{lem}\label{lem:basbip}
Let $G$ be a graph. Then:
\begin{itemize}
\item[(a)] $G$ is bipartite $\iff \hom(G,P_2)\ne 0$.
		
\item[(b)] If $G$ is connected and bipartite, then $\hom(G,P_2)= 2$.
		
\item[(c)] $G$ is bipartite $\iff \hom(G,H)\ne 0$ for all graphs $H$ with
    at least one edge.
		
\item[(d)] $G$ is bipartite $\iff$ $G$ does not contain a cycle of odd
	length.
		
\item[(e)] If $G$ is bipartite and $F$ is not, then $\hom(F,G)= 0$.
		
\item[(f)] If $G$ is bipartite, then $G$ is determined (up to isomorphism)
	by the values $\hom(F, G)$ for the bipartite graphs $F$ with $F\le G$
	\big(by the Lov\'asz Isomorphism Theorem and part~(e)\big).
\end{itemize}
\end{lem}

\noindent Let $G$ and $H$ be graphs. Then $G\times H$, the \emph{product of
$G$ and $H$} is the graph with $V(G\times H):= \{(u,v) \mid u\in V(G),\ v\in
V(H)\}$ and
\[
\big\{(u,v),(u',v')\big\}\in E(G\times H) \ \
\iff \ \ \{u,u'\}\in E(G) \text{ \ and\ } \ \{v,v'\}\in E(H).
\]
One easily verifies that for any graph $F$,
\begin{equation}\label{eq:homtimes}
\hom(F,G\times H)= \hom(F,G)\cdot \hom(F,H).
\end{equation}
Besides the simple facts on bipartite graphs mentioned above, we also need a
deep result.

\begin{theo}[\LV\ Cancellation Law~\cite{lov71}]\label{thm:canclaw}
Let $H$ be a graph. Then
\[
\text{$H$ is not bipartite $\iff$ for all graphs $F$ and $G$,
\big($F\times H\cong G\times H$ implies $F\cong G$\big).}
\]
\end{theo}

\noindent
The following lemma contains a further step for the proof of
Theorem~\ref{thm:3adapiso}.
\begin{lem}\label{lem:finF}
Let $n\ge 2$ and let $t$ be the smallest natural number with $n\le 2t$. We
set (recall that $C_m$ denotes a cycle of length $m$)
\[
\cls K_t :=
\big\{F \bigmid \text{$\hom(F,C_{2t+1})> 0$ \ and \ $|V(F)|\le (2t+1)^2$}\big\}.
\]
Let $G$ and $H$ be graphs with $|V(G)|=|V(H)|=n$ and $\hom(F,G)=\hom(F,H)$
for all $F\in\cls K_t$. Then $G\cong H$.
\end{lem}

\proof For a contradiction, assume $G\not\cong H$. The cycle $C_{2t+1}$ is
not bipartite (see Lemma~\ref{lem:basbip}(d)). Thus, by the \LV\ Cancellation
Law,
 \[
G\times C_{2t+1} \not\cong H \times C_{2t+1}.
\]
As $G\times C_{2t+1}$ has $n\cdot(2t+1)$ vertices, by the \LV\ Isomorphism
Theorem (see Theorem~\ref{thm:loviso}) there is a graph $F$ with $|V(F)|\le
n\cdot(2t+1)\le (2t+1)^2$ such that
\begin{equation}\label{eq:finF1}
\hom(F, G \times C_{2t+1})\ne \hom(F, H \times C_{2t+1}).
\end{equation}
If $F \notin\cls K_t$, then $\hom(F,C_{2t+1})= 0$ and thus
by~\eqref{eq:homtimes}, $\hom(F, G \times C_{2t+1})=0= \hom(F, H \times
C_{2t+1})$, a contradiction to~\eqref{eq:finF1}. Hence, $F\in \cls K_t$ and
so by assumption, $\hom(F,G)= \hom(F,H)$. However, by~\eqref{eq:homtimes}
this again contradicts~\eqref{eq:finF1}. \proofend

\medskip
\noindent \textit{Proof of Theorem~\ref{thm:3adapiso}:}
The case $n= 1$ is trivial. So we assume $n\ge 2$.
Let again $t$ be the smallest natural number with $n\le 2t$ and
\[
\cls K_t:= \big\{F\bigmid \text{$\hom(F,C_{2t+1})> 0$ \ and \ $|V(F)|\le (2t+1)^2$}\big\}.
\]
For the class $\cls K_{\textup{bip}}$ of bipartite graphs in $\cls K_t$ let
$F_1\ (=F_1(n))$ be the graph constructed in Lemma~\ref{lem:encoding} for
$\cls K_{\textup{bip}}$ (i.e., $F_1= F_{\cls K_{\textup{bip}}}$). As $P_2\in
K_{\textup{bip}}$, we know that $E(F_1)\ne\emptyset$ and as the disjoint
union of bipartite graphs is bipartite, the proof of Lemma~\ref{lem:encoding}
shows that $F_1$ is bipartite. Let $F_2\ (=F_2(n))$ be the graph constructed
in Lemma~\ref{lem:encoding} for the class $\cls K_t\setminus \cls
K_{\textup{bip}}$.

To finish the proof of Theorem~\ref{thm:3adapiso} we have to show that for
graphs $G$ and $H$ with $|V(G)|= |V(H)|= n$,
\[
\text{$\hom(F_i, G)= \hom(F_i, H)$ for $i\in[2]$ \ implies \
	 $G\cong H$}.
\]
If $G$ and $H$ both have no edge, then clearly $G\cong H$. If, say $G$ has an
edge but $E(H)= \emptyset$, then, as $F_1$ is bipartite, $\hom(F_1, G)\ne
0=\hom(F_1, H)$ by Lemma~\ref{lem:basbip}~(c) and $E(F_1)\ne \emptyset$.
Hence we can assume that both graphs contain at least one edge.

For a contradiction assume that
\begin{equation}\label{eq:neukon}
\text{$\hom(F_i, G) = \hom(F_i, H)$ for $i\in[2]$ \ and \ $G\not\cong H$.}
\end{equation}
Then, by Lemma~\ref{lem:finF} there is a graph $F_0\in \cls K_t$ with
\begin{equation}\label{eq:eqnot}
\hom(F_0, G)\ne \hom(F_0, H).
\end{equation}
Assume first that $F_0\in \cls K_{\textup{bip}}$. As $G$ and $H$ contain at
least one edge, by Lemma~\ref{lem:basbip}~(c)
\begin{eqnarray*}
\hom(F, G)> 0 & \text{and} & \hom(F, H) > 0,
\end{eqnarray*}
for every bipartite graph $F$. In particular, this holds for all graphs $F$
in $\cls K_{\textup{bip}}$, i.e., for $\cls K:=\cls K_{\textup{bip}}$ part
(a) in Lemma~\ref{lem:encoding} is not true. Thus, as $\hom(F_1, G)=
\hom(F_1, H)$ (by~\eqref{eq:neukon}), we know that part (b) in
Lemma~\ref{lem:encoding} holds (recall that $F_1= F_{\cls
K_{\textup{bip}}}$). I.e., for every $F \in \cls K_{\textup{bip}}$,
\[
\hom(F, G) = \hom(F, H).
\]
In particular, this holds for $F= F_0$ contradicting~\eqref{eq:eqnot}.

Thus $F_0\in \cls K_t\setminus\cls K_{\textup{bip}}$. Then $\hom(F_0, F)=0$
for all bipartite graphs~$F$ \big(by Lemma~\ref{lem:basbip}(e)\big). Hence,
by~\eqref{eq:eqnot} at least one of $G$ and $H$ must be non-bipartite. By
$\hom(F_2, G)= \hom(F_2, H)$, Lemma~\ref{lem:encoding} for $\cls K:= \cls
K_t\setminus\cls K_{\textup{bip}}$, and~\eqref{eq:eqnot} there exist graphs
$F^G$ and $F^H$ in $\cls K_t\setminus \cls K_{\textup{bip}}$ with
\[
\hom(F^G, G) = \hom(F^H, H) = 0.
\]
Without loss of generality, suppose that $G$ is not bipartite and thus
contains an odd cycle, say of length~$\ell$. Since $\ell\le n\le 2t+1$, we
have
\[
\hom(C_{2t+1}, C_\ell)> 0.
\]
In fact, one easily verifies that $\hom(C_{k}, C_m)> 0$ for odd $k$ and $m$
with $k>m$. As $F^G\in \cls K_t$, we know that $\hom(F^G,C_{2t+1})>0$.
Therefore, $\hom(F^G,C_{\ell})>0$, which implies $\hom(F^G,G)>0$, a
contradiction. \proofend

As already mentioned two adaptive hom-queries cannot determine the
isomorphism type of a graph. Moreover, even some simple properties cannot be
decided by a 2 adaptive hom algorithm:

\begin{theo}\label{thm:twonot}
The class of cycles has no 2 adaptive hom algorithm.
\end{theo}

\proof We only need to show that for every graph $F$ and function $N: \mathbb
N\to \textsc{Graph}$, there exist some graphs $G$ and $H$ such that
\begin{itemize}
\item[(a)] \text{$n_0: =\hom(F, G) = \hom(F, H)$ and $\hom(N(n_0), G) =
    \hom(N(n_0), H)$}.

\item[(b)] $G$ is a cycle and $H$ is not.
\end{itemize}
For every such $F$ and $N$, let $k:= |V(F)|$ and choose $\ell\in \mathbb N$ with $4\ell+
2> k$. We consider the following three non-isomorphic graphs:
\begin{itemize}
\item the bipartite graph $G:= C_{12\ell+ 6}$,
	
\item the bipartite graph $H_1:= C_{4\ell+ 2}\dotcup C_{4\ell +2} \dotcup C_{4\ell+2}$.

\item and the non-bipartite graph $H_2:= C_{6\ell+ 3} \dotcup C_{6\ell+ 3}$,

\end{itemize}

\bigskip \noindent \textit{Claim.}
$\hom(F, G)= \hom(F,H_1)= \hom(F,H_2)$.

\medskip
\noindent \textit{Proof of Claim:} If $F$ is non-bipartite, then
$\hom(F, G)= \hom(F, H_1)= 0$ \big(by
Lemma~\ref{lem:basbip}~(e)\big). Any odd cycle in $F$ \big(at least
one exists by Lemma~\ref{lem:basbip} (d)\big) has length at most
$|V(F)|= k < 4\ell+ 2< 6\ell + 3$. Thus there is no homomorphism from
$F$ to $C_{6\ell+ 3}$. It implies $\hom(F, H_2)= 0$ as well.

\smallskip
Assume that $F$ is a bipartite graph. Let $A_1, A_2, \ldots, A_p$ be
the connected components of $F$. Note that each $A_i$ has at most $k$
vertices. We fix a vertex $v_i$ in each $V(A_i)$. For $t\ge 3$ consider a
cycle~$C_t$ and again fix an arbitrary $u_t\in V(C_t)$. We set
\[
\rho(i, t):= \big|\{h\in \Hom(A_i, C_t)\mid h(v_i)= u_t\}\big|.
\]
Clearly,
\[
\hom(A_i, C_t)= t\cdot \rho(i, t).
\]
The key observation is that for $t,t'> k$
\[
\rho(i, t)= \rho(i, t').
\]
This follows easily from the fact that each $A_i$ is connected and has at
most $k= |V(F)|$ vertices and thus, for $h\in \Hom(A_i, C_t)$ with $h(v_i)=
u_t$ every vertex in the image of $h$ has distance less than $k$ from $u_t$.
Hence
\[
\hom(A_i, G)=\hom(A_i, H_1)=\hom(A_i, H_2)=(12\ell + 6)\cdot \rho(i, k+1).
\]
Then, by~\eqref{eq:unionhom},
\[
\hom(F, G)=\hom(F, H_1)=\hom(F, H_2)=
\prod_{i\in [p]} (12\ell + 6)\cdot \rho(i, k+1).
\]
\hfill $\dashv$

\medskip
\noindent Let $n_0:= \hom(F, G)= \hom(F, H_1)= \hom(F, H_2)$. Moreover, let
\[
F^*:= N(n_0).
\]
Assume first that $F^*$ is non-bipartite. Since both $G$ and $H_1$ are
bipartite, we conclude \big(by Lemma~\ref{lem:basbip}~(e)\big)
\[
\hom(F^*, G)= \hom(F^*, H_1)=0.
\]
We find the required graphs for $G:= G$ and $H:=H_1$. Thus, $F^*$ is
bipartite. We show
\[
\hom(F^*, G)= \hom(F^*, H_2).
\]
Then we can find the required graphs for $G:= G$ and $H:= H_2$.

\smallskip
It is straightforward to verify that $G\cong P_2\times C_{6\ell+3}$. Recall
that $H_2= C_{6\ell+3}\dotcup C_{6\ell+3}$. Assume that $F^*$ has the
connected components $A_1, \ldots, A_p$. Since each $A_i$ is bipartite and
connected, we have $\hom(A_i, P_2)= 2$ \big(by
Lemma~\ref{lem:basbip}~(b)\big). Then
\begin{align*}
 \hom(F^*, H_2)& = \prod_{i\in [p]} \hom(A_i, C_{6\ell+3}\dotcup C_{6\ell+3})
   & \text{(by \eqref{eq:unionhom})} \\
 & = \prod_{i\in [p]} 2\cdot \hom(A_i, C_{6\ell+3})
   & \text{(as the $A_i$'s are connected)}\\
 & = \prod_{i\in [p]} \hom(A_i, P_2)\cdot \hom(A_i, C_{6\ell+3}) \\
 & = \prod_{i\in [p]} \hom(A_i, P_2 \times C_{6\ell+3})
   & \text{(by \eqref{eq:homtimes})} \\
 & = \prod_{i\in [p]} \hom(A_i, G) =\hom(F^*, G)
   & \text{\Big(by $G\cong P_2 \times C_{6\ell+3}$ and $F^* =\dot{\bigcup}_{i\in [p]}A_i$\Big)}.
\end{align*}
This finishes our proof.
\proofend

\begin{cor}\label{cor:twonot}
There is no graph $F$ and no function $N: \mathbb N\to \textsc{Graph}$ such
that for all graphs $G$ and~$H$,
\begin{align}\label{eq:two}
\text{if\/ $n_0: =\hom(F, G) = \hom(F, H)$ and
 $\hom(N(n_0), G) = \hom(N(n_0), H)$, then $G\cong H$}.
\end{align}
\end{cor}

\section{Right hom algorithms}\label{sec:right}

In~1993 Chaudhuri and Vardi~\cite{chavar93} (see also~\cite{atskolwu21})
showed the analogue of the \LV\ Isomorphism Theorem for the right
homomorphism vector of a graph $G$. More precisely, the vector of values
$\hom(G,F)$, where $F$ ranges over all graphs, characterizes the isomorphism
type of~$G$. In this section we will see that our main ``positive'' results
on hom algorithms fail for right hom algorithms (see
Proposition~\ref{pro:fails}) and that various results that survive use a
completely different proof technique. Note that a graph $G$ is 3-colorable if
$\hom(G,K_3)> 0$ in contrast to Theorem~\ref{thm:plan}.


\begin{defn}\label{def:rhom}
A class $\cls C$ of graphs can be \emph{decided by a right hom algorithm} if
and only if there is a $k\ge 1$ and graphs $F_1, \ldots, F_k$ such that for
all graphs $G$ and $H$,
\[
\text{$\hom(G,F_i)=\hom(H,F_i)$ for all $i\in[k]$ \ \
 implies \ \ $(G\in\cls C\iff H\in\cls C)$},
\]
or equivalently, if in addition to $F_1, \ldots, F_k$ there is a set
$X\subseteq \mathbb N^k$ such that for any graph $G$, \big($G\in \cls C\iff
(\hom(G,F_1),\ldots, \hom(G,F_k))\in X$\big).
\end{defn}
Again one can show that $\cls C$ is decidable if and only if a decidable set
$X$ can be chosen:

\begin{prop}\label{pro:rxc}
Let $\cls C$ have a right hom algorithm based on $F_1, \ldots, F_k$. Then the
set $X$ can be chosen to be decidable if and only if $\cls C$ is decidable.
\end{prop}

\proof
We set $X(\cls C):=\{\big(\hom(G,F_1),\ldots, \hom(G,F_k)\big) \mid G\in\cls
C\}$. Then, it suffices to show that ($\cls C$ is decidable if and only if
$X(\cls C)$ is decidable).

Again the direction from right to left is trivial. Let $\cls F:=\{F_1,\ldots,
F_k\}$. It is easy to see that the other direction follows from the following
claim. We set
\[
\cls F:=\{F_1,\ldots,
F_k\}
\]
\noindent \textit{Claim.} Let $G$ be a graph. Then there is a graph $\bar G$
with $\hom(\bar G,F)= \hom(G,F)$ for all $F\in\cls F$ and
\begin{itemize}
\item if $\cls F=\{F \}$ and $F$ only contains a single vertex and $G$
    contains at least an edge, then $|V(\bar G)|\le 2$;

\item if $\hom(G,F)= 0$ for all $F\in\cls F $, then $|V(\bar G)|\le
    1+\max\{|V(F)| \mid F\in \cls F \}$;

\item otherwise, $|V(\bar G)|\le \prod_{F\in\cls F}|V(F)|^{\hom(G,F)}$.
\end{itemize}

\noindent \textit{Proof of Claim.} In the first case we set
 set $\bar G:=P_2$.
In the second case let
$\bar G$ be a clique of size $1+ \max\{|V(F)| \mid F\in \cls F \}$.

Otherwise we define the following equivalence relation $\sim$ on $V(G)$,
\begin{eqnarray*}
u\sim v & \iff &
\text{for all $F\in \cls F$ and all $f\in \Hom(G,F): \ f(u)= f(v)$}.
\end{eqnarray*}
Denote by $\bar u$ the equivalence class of $u$. Define the graph $\bar
G=(\bar V, \bar E)$ by $\bar V:= \{\bar u \mid u\in V(G)\}$ and
\begin{equation}\label{eq:defbarg}
\bar E:=\{\bar u\bar v
 \mid \text{$\bar u\ne \bar v$
 and there are $u'\in \bar u$ and $v'\in \bar v$ with $u'v'\in E(G)$}\}.
\end{equation}
Then, as every $f\in \Hom(G,F)$ for some $F\in\cls F$ ``leads'' to at most
$|V(F)|$ equivalence classes, we get:
\[
|V(\bar G)|\le \prod_{F\in\cls F}|V(F)|^{\hom(G,F)}.
\]
Now a straightforward proof shows $\hom(\bar G,F)=\hom(G,F)$ for all $F\in
\cls F$.
\proofend

It should be clear how we define \emph{$k$ adaptive right hom
algorithms} for a class $\cls C$ of graphs: we just replace in
Definition~\ref{def:ada}
\[
\text{$n_1:= \hom(F, G)$ \ \ and \ \ } n_2:= \hom(N(n_1), G),\ \
\ldots, \ \ n_k:= \hom(N(n_1, n_2, \ldots, n_{k-1}), G)
\]
by
\begin{equation}\label{eq:radar}
\text{$n_1:= \hom(G, F)$ \ \ and \ \ } n_2:= \hom(G, N(n_1)),\ \
\ldots, \ \ n_k:= \hom(G,N(n_1, n_2, \ldots, n_{k-1}))
\end{equation}

\smallskip
The failure in the ``right world'' of the ``positive'' results
Theorem~\ref{thm:exiuni} and Theorem~\ref{thm:3adapsuf} is shown by:

\begin{prop}\label{pro:fails}
Let $k\ge 1$. The class $\cls K(3)$ of graphs containing a clique of size $3$
(expressible by the existential sentence $\exists x \exists y \exists
z(Exy\wedge Eyz\wedge Exz)$) cannot be decided by a $k$ adaptive right hom
algorithm (and hence not by a right hom algorithm).
\end{prop}

\proof
For a graph $G$ we denote by $\chi(G)$ the chromatic number of $G$, i.e., the
least $s$ such that~$G$ is $s$-colorable. Clearly, $\big(m< \chi(G)\iff
\hom(G,K_m)= 0 \big)$ for the clique~$K_m$ with~$m$ elements, and hence, for
every graph $F$,
\begin{equation}\label{eq:chro}
\text{if $|V(F)|< \chi(G)$, then $\hom(G,F)= 0$.}
\end{equation}
For a contradiction, assume that $F$, $N$, and $X$ (compare
Definition~\ref{def:ada} and \eqref{eq:radar}) witness the existence of a $k$
adaptive right hom algorithm for~$\cls C$. Let $s> 3$ be bigger than
\[
 \max\{|F|, |N(0)|, |N(0,0)|,\ldots, |N(\underbrace{0,\ldots, 0}_{k-1\ \textup{times}})| \}
\]
According
to~\cite{myc08} there is a $G_0\notin \cls K(3)$ such that $\chi(G_0)= s$.
Thus by~\eqref{eq:chro}, we have
\[
\hom(G_0, F)= 0, \hom(G_0, N(0))=0, \hom(G_0, N(0,0))=0\, \ldots, \hom(G_0,N(\underbrace{0,\ldots, 0}_{k-1\ \textup{times}})
= 0
\]
 and hence, $(0,0,\ldots, 0)\notin X$. However,
 $K_s\in\cls K(3)$ and
\[
\hom(K_s, F)= 0, \hom(K_s, N(0))=0, \hom(K_s, N(0,0))=0\, \ldots, \hom(K_s,N(\underbrace{0,\ldots, 0}_{k-1\ \textup{times}})
= 0
\]
and thus $(0,0,\ldots, 0)\in X$, a contradiction.
\proofend
We mention a positive result on right hom algorithms.

\begin{theo}\label{thm:finedg}
Every class $\cls C$ of graphs with the property that there is a bound on the
number of edges of graphs in $\cls C$ has a right hom algorithm.
\end{theo}
We get the following corollaries:
\begin{cor}\label{cor:fns}
A class of graphs containing only finitely many graphs has a right hom
algorithm.
\end{cor}

\begin{cor}\label{cor:ner}
If $\varphi$ is an \FO-sentence not containing the edge relation symbol, then
the class $\cls C(\varphi)$ of graphs models of $\varphi$ has a right hom
algorithm.
\end{cor}
Note that for such a class $\cls C(\varphi)$ there is a finite or cofinite
set $Z$ of positive integers such that a graph $G$ is in $\cls C(\varphi)$ if
and only if $|V(G)| \in Z$. One gets this corollary by passing from the class
$\cls C(\varphi)$ to the class obtained by deleting all edges in the graphs
of $\cls{\varphi}$ and applying Theorem~\ref{thm:finedg}.
\medskip

The following lemma, which already yields the last corollary, is the main
tool of a proof of Theorem~\ref{thm:finedg}. Thereby we use the fact that for
graphs $G$ and $H$,
\begin{equation}\label{eq:rhle}
\text{$G\cong H \iff \hom(G,F)=\hom(H,F)$
 for all $F$ with $|V(F)|\le \max\{|V(G)|,|V(H)|\}$}
\end{equation}
(e.g., it can be obtained by Theorem~\ref{thm:loviso} from \cite[Theorem
3]{atskolwu21} by taking as $\mathscr F$ the class of graphs with at most
$\max\{|V(G)|,|V(H)|\}$ vertices).
\begin{lem}\label{lem:powright}
Let $k\ge 2$ and $\cls F(k)$ be the class of graphs with at most $k^3$
vertices. Then there are no graph $G$ and $H$ such that
\[
\text{$\hom(G,F)=\hom(H,F)$ for all $F \in \cls F(k)$ and $|V(G)|\le k$ and $|V(H)| > k$}.
\]
\end{lem}

\proof
For a contradiction suppose that $G$ and $H$ are such graphs. Then,
by~\eqref{eq:rhle}, $|V(H)|>k^3$. As $|V(G)|\le k$, we have $\hom(G, K_{k^3})\le
k^{3k}$.
 	
On the other hand, $\hom(G, K_k)> 0$ by $|V(G)|\le k$. As $\hom(H, K_k)=
\hom(G, K_k)$, we get $\hom(H, K_k)> 0$, i.e., $H$ is $k$-colorable. Let $c:
V(H)\to [k]$ be a $k$-coloring of $H$. For every $i\in [k]$ we set
\[
V_i:= \{v\in V(H)\mid c(v)=i\}.
\]
As $|V(H)|\ge k^3$, without loss of generality we can assume that
$\big|V_1\big|\ge k^3/k\ge k^2$. In particular, we can write $V_1= \{u_1,
u_2, \ldots, u_{k^2}, \ldots \}$ with pairwise distinct $u_i$'s. Now consider
functions $c': V(H)\to [k^3]$ satisfying
\begin{align*}
& c'(u) = c(u)\ (\in [2,k]) & \text{for $u\notin V_1$} \\
& c'(u_1) = c(u) = 1 \\
& c'(u_i)> k & \text{for $i\ge 2$}.
\end{align*}
It is easy to check that such a $c'$ is a $k^3$-coloring of $H$. Moreover,
the number of such $c'$ is
\[
(k^3-k)^{|V_1|-1}\ge (k^3-k)^{k^2-1}.
\]
Putting all pieces together, we get
\begin{align*}
\hom(H, K_{k^3}) = \text{number of $k^3$-colorings of $H$}
\ge (k^3-k)^{k^2-1}> (k^2)^{k^2-1} \ge k^{3k}
\end{align*}
for $k\ge 2$. But we have seen above that $k^{3k}\ge \hom(G, K_{k^3})$.
Hence $\hom(H, K_{k^3})\ne \hom(G, K_{k^3})$, the desired contradiction.
\proofend

\noindent \textit{Proof of Theorem~\ref{thm:finedg}:} For a graph $G$ we
denote by $G^\wi$ (``$G$ without isolated vertices'') the graph obtained from
$G$ by deleting the isolated vertices and let $i(G)$ be the number of
isolated vertices of $G$. Note that for any graph $F$,
\begin{equation}\label{eq:hwi}
\hom(G,F)= \hom(G^\wi,F)\cdot |V(F)|^{i(G)}.
\end{equation}
Let $k$ be an upper bound for the number of edges in graphs of $\cls C$.
Hence $|G^\wi|\le 2k$ for $G\in \cls C$. By Lemma~\ref{lem:powright} we know
that every graph with at most $2k$ elements can be separated from graphs with
more than $2k$ vertices by the class of graphs in $\cls F_{2k}$, the class of
graphs with at most $(2k)^3$ vertices and hence by a set $\cls F^0_{2k}$ that
contains one graph of every isomorphism type of graphs with at most $(2k)^3$
vertices Let
\begin{equation}\label{eq:efphi}
\cls F_{\cls C}:= \cls F^0_{2k}\cup \{F\dotcup K_1 \mid F\in \cls F^0_{2k} \}.
\end{equation}
We show that $\cls F_{\cls C}$ are the graphs of a right hom algorithm for
$\cls C$. For a contradiction assume that $G\in\cls C$ and $H\notin \cls C$
and
\begin{equation}\label{eq:efphi}
\text{ for all $F\in\cls F_{\cls C}$: \quad $\hom(G,F)= \hom(H,F)$}.
\end{equation}
Hence, by \eqref{eq:hwi},
\[
\text{ for all $F\in\cls F_{\cls C}$: \quad $\hom(G^\wi,F)\cdot |V(F)|^{i(G)}= \hom(H^\wi,F)\cdot |V(F)|^{i(H)}$}.
\]
i.e.,
\begin{equation}\label{eq:efphi}
\text{ for all $F\in\cls F_{\cls C}$: \quad $\hom(H^\wi,F)=\hom(G^\wi,F)\cdot |V(F)|^{i(G)-i(H)}$}.
\end{equation}
In particular,
\[
\hom(G^\wi,F)=0\iff \hom(H^\wi,F)=0.
\]
Let $F_0 \in\cls F_{\cls C}$ be such that $\hom(G^\wi,F_0)\ne 0$. As
$\hom(G^\wi,F_0\dotcup K_1)= \hom(G^\wi,F_0)$, we can assume that $F_0\dotcup
K_1 \in\cls F_{\cls C}$ (see the definition of $\cls F_{\cls C}$). Thus, by
\eqref{eq:efphi} for $F:=F_0\dotcup K_1$,
\[
\hom(H^\wi,F_0\dotcup K_1)=\hom(G^\wi,F_0\dotcup K_1)\cdot |V(F_0\dotcup K_1)|^{i(G)-i(H)},
\]
i.e.,
\[
\hom(H^\wi,F_0)=\hom(G^\wi,F_0)\cdot (|V(F_0)|+1)^{i(G)-i(H)}.
\]
Together with \eqref{eq:efphi} for $F:=F_0$, this implies
$|V(F_0)|^{i(G)-i(H)}= (|V(F_0)|+1)^{i(G)-i(H)}$ and thus, $i(G)=i(H)$.
Altogether we have
\begin{equation}\label{eq:schlu}
\hom(G^\wi,F)=\hom(H^\wi,F) \text{\ \ for all $F\in\cls F_{\cls C}$. }.
\end{equation}
As $|V(G^\wi)|\le 2k$ and $\cls F^0_{2k}\subseteq F_{\cls C} $, we get
$|V(H^\wi)|\le 2k$. Therefore, by~\eqref{eq:schlu}, $G^\wi=H^\wi$ and hence,
$G\cong H$, a contradiction.\proofend

We should remark that some results with non-trivial proofs for hom algorithms
are trivial for right hom algorithms. For example, the following simple proof
shows that there is no right hom algorithm for the class $\cls
C(\forall x\exists y Exy)$ of graphs with at least one isolated vertex.

Let $F_1,\ldots, F_k$ be any finite set of graphs and set $m:=
1+\max\{|V(F_i)| \mid i\in [k]\}$. Then, $\hom(K_m,F_i)= 0= \hom(K_m\dotcup
K_1)$ for every $i\in [k]$. Thus there is no right hom algorithm that detects
an isolated vertex.

\section{Conclusions}

To the best of our knowledge this is the first paper that systematically
analyzes properties of graphs that can be decided by a constant number of
homomorphism counts. We gain a quite satisfactory picture. Separately we
consider non-adaptive hom algorithms and adaptive hom algorithms.
We characterize those prefix classes of first-order logic with the property
that all classes of graphs definable by a corresponding first-order sentence
have a non-adaptive query-algorithm with a constant number of homomorphism
counts. Furthermore, we show that every class of graphs can be recognized by
an adaptive hom algorithm with three homomorphism counts. We present an
example where two counts are not sufficient. In general, given a class of
graphs, the three adaptive hom algorithm we get for this class needs
for a graph $G$ homomorphism counts $\hom(F,G)$ where the size of $F$ is
superpolynmial in the size of~$G$. We believe that this is necessary for some
classes. It is a challenging task for graph classes that are relevant in
applications to analyze the existence of an algorithm where the size of the
corresponding~$F$'s are polynomial in the size of $G$.

\medskip
\subsection*{Acknowledgement}
The collaboration of the first two authors is funded by the Sino-German
Center for Research Promotion (GZ 1518). Yijia Chen is also supported by
National Natural Science Foundation of China (Project 61872092).

\bibliographystyle{plain}
\bibliography{refer}

\end{document}